\documentclass[12pt]{iopart}

\usepackage[utf8]{inputenc} 
\usepackage[T1]{fontenc}
\expandafter\let\csname equation*\endcsname\relax
\expandafter\let\csname endequation*\endcsname\relax
\usepackage{amsmath}
\usepackage{hyperref} 
\usepackage{cleveref} 
\hypersetup{colorlinks = true, 
	      linkcolor = black, 
	      urlcolor = black,
            citecolor = blue} 
\usepackage{url}            
\usepackage{booktabs} 
\usepackage{amsfonts}       
\usepackage{nicefrac}      
\usepackage{microtype}           
\usepackage[pdftex]{graphicx}       
\usepackage{subfigure}
\usepackage{color}
\usepackage{qcircuit}
\usepackage{float} 
\usepackage{amsthm}
\usepackage{cite}

\usepackage{glossaries}

\newacronym{gcd}{GCD}{Greatest Common Divisor}
\newacronym{lcm}{LCM}{Least Common Multiple}

\newacronym{qec}{QEC}{Quantum Error Correction}
\newacronym{nisq}{NISQ}{Noisy Intermediate-Scale Quantum}
\newacronym{hdrg}{HDRG}{Hard Decision Renormalisation Group}
\newacronym{rg}{RG}{Renormalisation Group}
\newacronym{ftqc}{FTQC}{Fault-Tolerant Quantum Computation}
\newacronym{spam}{SPAM}{State Preparation and Measurement}
\newacronym{mwpm}{MWPM}{Minimum Weight Perfect Matching}

\newacronym{ler}{LER}{Logical Error Rate}

\newacronym{ann}{ANN}{Artificial Neural Network}

\newacronym{fcnn}{FCNN}{Fully Convolutional Neural Network}
\newacronym{cnn}{CNN}{Convolutional Neural Network}
\newacronym{lstm}{LSTM}{Long Short-Term Memory}
\newacronym{gan}{GAN}{Generative Adversarial Network}
\newacronym{ml}{ML}{Machine Learning}

\newacronym{uf}{UF}{Union-Find}

\newacronym{pcm}{PCM}{Parity Check Matrix}

\newacronym{dem}{DEM}{Detector Error Model}

\newacronym{fpga}{FPGA}{Field Programmable Gate Array}
\newacronym{asic}{ASIC}{Application-Specific Integrated Circuit}

\newacronym{cafe}{CAFE}{Context Aware Fidelity Estimation}

\newacronym{sl}{SL}{Supervised Learning}
\newacronym{rl}{RL}{Reinforcement Learning}
\newacronym{rnn}{RNN}{Recurrent Neural Network}
\newacronym{ffnn}{FFNN}{Feed Forward Neural Network}
\newacronym{rbm}{RBM}{Restricted Boltzmann Machine}

\usepackage{lipsum}

\graphicspath{{media/}}

\usepackage{iopams} 
\begin{document}

\newcommand{\paperthreetitle}{Fully convolutional 3D neural network decoders for surface codes with syndrome circuit noise}
\newcommand{\paperthreetitleshort}{ }

\newcommand{\etaldot}{\textit{et al.}}

\title[\paperthreetitleshort ]{\paperthreetitle }

\author{Spiro Gicev$^{1,2*}$, Lloyd C. L. Hollenberg$^{1,2}$ and Muhammad Usman$^{2,3}$}
\address{$^1$ Center for Quantum Computation and Communication Technology, School of Physics, University of Melbourne, Parkville,
3010, VIC, Australia}
\address{$^2$ School of Physics, University of Melbourne, Parkville,
3010, VIC, Australia}
\address{$^3$ Data61, CSIRO, Clayton, VIC 3168, Australia}
\address{$^*$ Author to whom any correspondence should be addressed.}
\ead{sgicev@student.unimelb.edu.au}

\vspace{10pt}

\begin{abstract}
\glspl{ann} are a promising approach to the decoding problem of \gls{qec}, but have observed consistent difficulty when generalising performance to larger \gls{qec} codes. Recent scalability-focused approaches have split the decoding workload by using local \glspl{ann} to perform initial syndrome processing and leaving final processing to a global residual decoder. We investigated \gls{ann} surface code decoding under a scheme exploiting the spatiotemporal structure of syndrome data. In particular, we present a vectorised method for surface code data simulation and benchmark decoding performance when such data defines a multi-label classification problem and generative modelling problem for rotated surface codes with circuit noise after each gate and idle timestep. Performance was found to generalise to rotated surface codes of sizes up to $d=97$, with depolarisation parameter thresholds of up to $0.7\%$ achieved, competitive with \gls{mwpm}. Improved latencies, compared with \gls{mwpm} alone, were found starting at code distances of $d=33$ and $d=89$ under noise models above and below threshold respectively. These results suggest promising prospects for \gls{ann}-based frameworks for surface code decoding with performance sufficient to support the demands expected from fault-tolerant resource estimates.
\end{abstract}

\vspace{2pc}

\maketitle


%
%
%
%

\section{Introduction}\label{Ch5-Intro}
\gls{qec} is expected to facilitate quantum speed-ups by implementing fault-tolerant quantum circuits on noisy hardware~\cite{PhysRevA.52.R2493, 548464}. Resource estimates for fault-tolerant algorithm implementations make assumptions regarding multiple characteristics of the underlying computational devices~\cite{Gidney_2021, doi:10.1073/pnas.1619152114}. One assumption involves the characteristics of quantum device noise, often described using parametrisation by a uniform error rate, $p$, associated with each quantum gate. The threshold results yielded by this approach suggest the possibility of arbitrarily error suppression at realistic physical error rates \cite{wang2009threshold, fowler2012proof}. Multiple experimental devices have demonstrated quantum operations with error rates below threshold, with some performing small-scale logical circuits~\cite{erhard2021entangling, Bluvstein2024}. Resource estimates have introduced another assumption described as the decoding accuracy and reaction time, which together specify expected \glspl{ler} and decoding algorithm latencies, respectively. An architecture's decoding reaction time can be the limiting factor governing the maximum depth of fault-tolerant algorithms, even on hardware which is otherwise well below threshold~\cite{delfosse2023choose}, highlighting the importance of efficiency when developing classical algorithms to support noisy quantum hardware \cite{Battistel_2023}.

Existing approaches to decoding can be broadly categorised to be based on either look-up tables~\cite{das2021lilliputlightweightlowlatencylookuptable}, tensor networks~\cite{PhysRevA.90.032326_tensor_network_decoder}, matching~\cite{fowler2012towards}, clustering~\cite{delfosse2021almost} or \gls{ml}~\cite{torlai2017neural}. In terms of accuracy, look-up tables, tensor network and \gls{ml} techniques have demonstrated performance comparable to optimal maximum-likelihood decoding under known error models. When details of underlying error models are absent or limited, \gls{ml} approaches have been shown to be the best performing~\cite{Bausch2024}, however with optimality more difficult to show. Beyond accuracy, the practicality of a decoding algorithms additionally depends on their latency and scalability. Look-up tables are limited by the memory they require. Tensor network techniques are limited by increasing computational complexity associated with the entanglement associated with their bond-dimension. \gls{ml} methods are limited primarily by their training time. In each case the most accurate algorithms are limited to low distance \gls{qec} codes. This highlights the importance of the accuracy-latency/scalability trade-off~\cite{delfosse2023choose}, and motivates the use of efficient heuristic algorithms based on matching and clustering, such as \gls{mwpm} and \gls{uf}~\cite{fowler2012towards, delfosse2021almost}. However, care must still be taken to manage the large overhead such methods yet demand when put in practice~\cite{Skoric2023}.

\gls{ann}-based decoding has had growing interest for various codes, error models, and \gls{ann} designs \cite{blue2025machinelearningdecodingcircuitlevel, cao2025generativedecodingquantumerrorcorrecting, bödeker2025interpretabilityneuralnetworkdecoders, hu2025efficientuniversalneuralnetworkdecoder, 10993142, 11000145, Maan2025, 10901425, PhysRevApplied.22.044031, zhong2024advantagequantumneuralnetworks, Bordoni2023, wang2023transformerqecquantumerrorcorrection, 10619589, PhysRevResearch.6.L032004, cao2023qecgptdecodingquantumerrorcorrecting, PhysRevResearch.7.013029, PhysRevResearch.7.023181, egorov2023endequivariantneuraldecoder, Choukroun_Wolf_2024, overwater2022neural ,10821258, Yon_2025, zhang2023scalablefastprogrammableneural, ueno2022neoqecneuralnetworkenhanced}. Ni~\cite{Ni_2020} investigated scalable decoding of toric codes using \glspl{cnn} structured similarly to \gls{rg} decoders. Under the bit-flip noise model, this decoder achieved performance comparable with \gls{mwpm}. Meinerz \etaldot~\cite{PhysRevLett.128.080505} investigated the decoding performance of toric codes with depolarising noise up to distance 255, and with depolarising noise and syndrome measurement errors up to distance 63. \gls{ann} decoders were constructed using dense layers, and translated adjacent to nontrivial syndrome bits to calculate a set of preliminary data qubit corrections. Any remaining nontrivial syndrome bits were decoded with \gls{mwpm} or \gls{uf}. In \cite{Gicev2023scalablefast} a similar approach for unrotated surface codes was developed independently using a \gls{fcnn}. Boundary information was included in the input, allowing generalisation to codes of different shapes and sizes, up to $d=1025$. Chamberland \etaldot~\cite{Chamberland_2023} developed on these results by investigating \glspl{fcnn} decoding on rotated surface codes suffering circuit noise using 3D convolutional layers. Ueno \etaldot~\cite{ueno2022neoqecneuralnetworkenhanced} performed resource estimates for implementations on superconducting hardware. Further work is yet required to achieve satisfactory \gls{ann} decoding under constraints imposed by realistic hardware.

In this work, we investigate \gls{ann} approaches to decoding of rotated surface codes suffering circuit noise facilitated by a data representation utilising tiled unit cells. We focus on systems of large code distances ($d>13$) and errors throughout each step of syndrome measurement circuits. We find that the periodic representation we used enabled significant vectorisation in simulated data preparation and augmentation. The periodic data representation also highlighted an interpretation of low-level decoding as a highly periodic form of conditional distribution sampling. Three decoders are described corresponding to a supervised multi-label classifier trained on unmodified data (corresponding to randomly sampled circuit noise), a supervised multi-label decoder trained on augmented data, and a diffusion model trained on unmodified data.

The remainder of this work is as follows. In Section~\ref{subsec:Problem_Formulation} we discuss the rotated surface code memory decoding problem under circuit noise, and the demands imposed by experimental hardware. In Section~\ref{subsec:ch5_data_preparation} we discuss how surface code simulations can be represented in a periodic representation, facilitating parallel simulation and augmentation of decoding data. In Section~\ref{subsec:ch5_ann_approaches} we investigate the \gls{ler} performance of \gls{ann} decoders trained in the framework of a a multi-label classification problem and a conditional diffusion problem. In Section \ref{subsec:ch5_timing} we investigate the timing of multi-label classification decoders. In section \ref{subsec:compare_sc_ion} we briefly comment on the demands imposed by superconducting and trapped atom/ion hardware. Finally, in Section~\ref{sec:Discussion} we discuss our results, summarising implications and promising future directions for large-scale \gls{ann} decoders.

\section{Results}\label{Ch5-Results}
\subsection{Problem Formulation}\label{subsec:Problem_Formulation}
\subsubsection{Rotated Surface Codes}
\begin{figure*}[tbp]
    \centering
    \includegraphics[width=\textwidth]{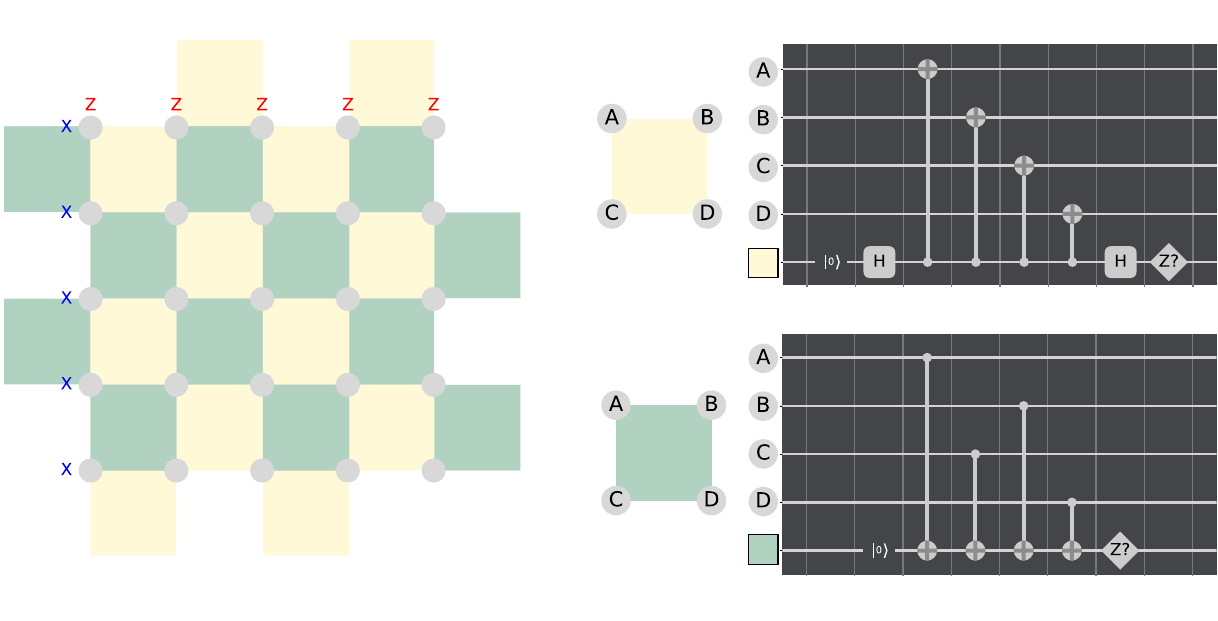}
    \caption{A distance 5 rotated surface code logical qubit with examples of an $X$ and $Z$ logical operator shown with chains of blue and red labels, respectively. Measurement circuits of $X$ and $Z$ surface code stabiliser operators are shown to the right. Data qubits are shown in grey. $X$ and $Z$ stabiliser operators are shown with yellow and green tiles, respectively.}
    \label{fig:sc_intro}
\end{figure*}

Surface codes are topological \gls{qec} codes compatible with square arrays of qubits with nearest neighbour connectivity. A distance 5 rotated surface code is shown in Figure~\ref{fig:sc_intro}. As stabiliser codes~\cite{10.1063/1.1499754}, surface codes use stabiliser operator measurements to extract information regarding locations of errors which may have occurred. Each stabiliser operator acts on up to four adjacent data qubits,
\begin{equation}
    V_i = \bigotimes_{j\in F_i} X_{j}, \quad P_i = \bigotimes_{j\in F_i}Z_{j},
\end{equation}
where $V_i$ and $P_i$ are vertex and plaquette operators associated with the face $F_i$, and $j\in F_i$ denotes the indices of data qubits on the border of $F_i$. Stabiliser generator operators near code boundaries act on fewer data qubits. Vertex and plaquette operators have $\pm 1$ eigenvalues and must simultaneously commute, restricting valid sets of stabilisers. A set of data qubits are in a surface code code-space state if and only if they are a simultaneous $+1$ eigenstate of all stabiliser generators. Operation which leaves all stabiliser eigenvalues unchanged correspond to logical operators. Nontrivial logical operators correspond to unbroken chains of data qubit operators between two distinct boundaries which commute with all stabilisers, as shown in Figure~\ref{fig:sc_intro}. While this formulation allows multiple distinct families of codes, rotated surface codes have recently dominated research interest and will be the focus of the remainder of this chapter.

Stabiliser operators may be measured simultaneously with the parallel application of two-qubit gates using ancilla qubits similar to those shown in Figure~\ref{fig:sc_intro}. One ancilla qubit is reserved for each stabiliser generator and two-qubit gate connectivity is required only between each ancilla and the data qubits the associated stabiliser generator acts upon. There is some flexibility in the gate sequences facilitating syndrome measurement. The most important consideration is to avoid single errors propagating to two-qubit errors in the same direction as logical operators~\cite{PhysRevA.90.062320}. The eight time step gate sequence used in this work contains modifications during the first and final measurement cycle involving data qubit reset and measurement, respectively. In this work, the optimisations associated with delayed reset and accelerated measurement of qubits near edges were neglected in order to keep all resets and measurements simultaneous during each cycle. As measurements proceed, decoding algorithms may be executed to find a set of errors consistent with the observed measurement values. This must be performed so that error corrected logical qubit measurements are known. While some decoders offer corrections to logical qubit measurements directly, the local relationship between errors and syndrome bits suggests the possibility of utilising significantly local approaches to decoding, and realising the associated computational benefits.

\subsubsection{Decoding Demands}
Rotated surface codes are expected to underpin quantum memory and logical computation. In these contexts, decoding must be performed at pace with the syndrome measurements of quantum hardware and with accuracy and scale sufficient to achieve the required suppression of \glspl{ler}. While additional nuances exist in precise resource estimation~\cite{10.1145/3624062.3624211, delfosse2023choose, Gidney2021howtofactorbit}, goals for decoding performance generally correspond to latency of approximately $1$ microsecond and \glspl{ler} of approximately $10^{-9}$ logical errors per $n_c=d$ syndrome measurement cycles. Surface code patches of size corresponding to distance $d=33$ logical qubits are usually the largest considered. Logical computation involves decoding systems of multiple connected surface code patches when lattice surgery is used, and multiple disconnected patches subject to correlated errors when transversal logical CNOT gates are used. The simulation and experimental progress towards logical operations of systems of increasing scales is an active area of research~\cite{Geher2024errorcorrected}.

\begin{figure*}[tbp]
    \centering
    \includegraphics[width=\textwidth]{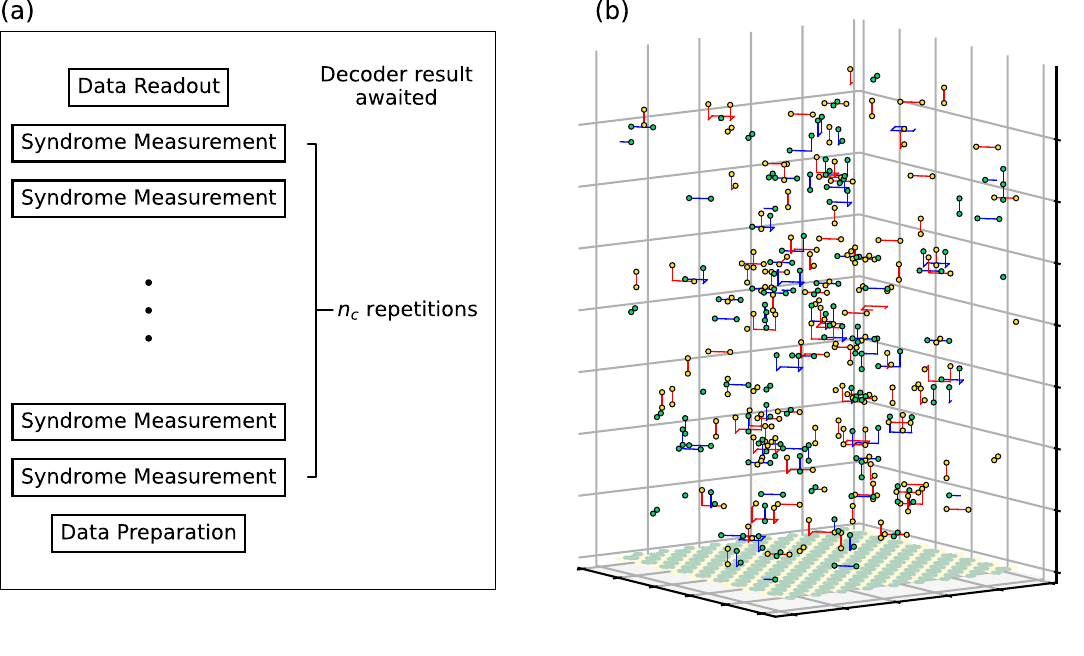}
    \caption{Visualising the decoding demands of surface codes. (a) shows a flowchart describing the classical processing demands of rotated surface codes during a quantum memory benchmark. (b) shows a space-time volume showing errors and syndrome changes associated with one quantum memory benchmark instance (time moves vertically upward). Blue and red lines are usually associated usually with $X$ and $Z$ data qubit errors, respectively. Vertical lines are associated with ancilla \gls{spam} errors. $X$ and $Z$ ancilla syndrome bit changes are yellow and green, respectively.}
    \label{fig:decoder_demands}
\end{figure*}

Quantum memory is a simpler benchmark which is regularly studied in simulation and experiment corresponding to initialisation, idling, and measurement of individual logical qubits prepared in the $X$ or $Z$ basis. A representation of a quantum memory benchmark is shown in Figure~\ref{fig:decoder_demands} (a). Data qubits are initially prepared in a particular product state associated with the desired logical qubit initialisation, with this information hidden until the evaluation step at the end of the benchmark. Syndrome measurement cycles then proceed, each consisting of time-steps corresponding to ancilla qubit initialisation, change of basis, CNOT gates, changing back to the $Z$ basis, and finally ancilla qubit measurement. Syndrome measurement cycles are repeated $n_c$ times to simulate behaviour when repeated cycles are performed to avoid logical time-like errors during logical qubit interactions. Finally, data qubits are measured, providing a final set of syndrome changes associated with the compatible basis as well as a logical qubit measurement result which awaits a possible correction. As many fault-tolerant logical operations produce additional feed-forward logical operations conditioned of measurement results, further calculation may need to be postponed until error corrected measurement results are known.

Figure~\ref{fig:decoder_demands} (b) shows a 3D representation of the classical data associated with a rotated surface code logical qubit during a quantum memory benchmark. Rounds where stabiliser generator measurement eigenvalues do not agree with the round prior are shown with coloured markers. The first round compares measured stabiliser values to the values associated with data qubit preparation. The final round corresponds to data qubit measurement, which is compared to the previous round of stabiliser measurement of the compatible basis. Calculating an appropriate logical correction can be performed by first prescribing a set of corrections compatible with the observed syndrome, and then calculating how these corrections would change the data qubit measurement values. In simulation, the rate at which errors and corrections result in unintended logical operations is used to quantitatively describe decoder performance. Alternatively, the rate at which final logical measurements disagree with data qubit preparation may also be used, and has the benefit of being able to be evaluated experimentally. The \gls{ler} requirements demanded by fault tolerant algorithm applications are usually set so the expected number of logical errors for the entire algorithm is less than one. From the higher-level perspective of a logical circuit with $n_{\mathrm{gates}}$ logical gates, this would demand the \gls{ler} $P_L$ per $n_c=d$ cycles defined by
\begin{equation}
    p_\mathrm{L}\propto\frac{1}{n_{\mathrm{gates}}}.
\end{equation}
From a lower level perspective, the \gls{ler} per $n_c=d$ cycles (where $d$ is the code distance) would need to be inversely proportional to the number of logical blocks, $n_{\mathrm{blocks}}$, defined by
\begin{equation}\label{Eq:blocks_LER}
    p_\mathrm{L}\propto\frac{1}{n_{\mathrm{blocks}}}.
\end{equation}
Codes are generally desired offering the property,
\begin{equation}\label{Eq:approx_LER}
    p_L \propto p^{d+1/2},
\end{equation}
which when combined with proportionality in Equation~\ref{Eq:blocks_LER}, code distance requirements grow slowly as
\begin{equation}
    \frac{d+1}{2}\propto \frac{\log(n_{\mathrm{blocks}})}{\log(1/p)},
\end{equation}
so that for a constant error rate, polynomially increasing the number of logical blocks can have errors sufficiently suppressed by linearly increasing code distances. It should be noted that higher order terms do exist in the true form of Equation~\ref{Eq:approx_LER}, and that the coefficients of the first few terms can be such that multiple terms are significant at error rates of interest.

Small variations in the above result will also exist due to edge effects and the variance of \gls{ler} between different blocks. Monte-Carlo simulations can be used to find accurate estimates for more accurate resource requirement estimates.

\subsection{Data Preparation}\label{subsec:ch5_data_preparation}
\subsubsection{Unit Cell Representation}
\begin{figure}[t]
    \centering
    \includegraphics[width=\columnwidth]{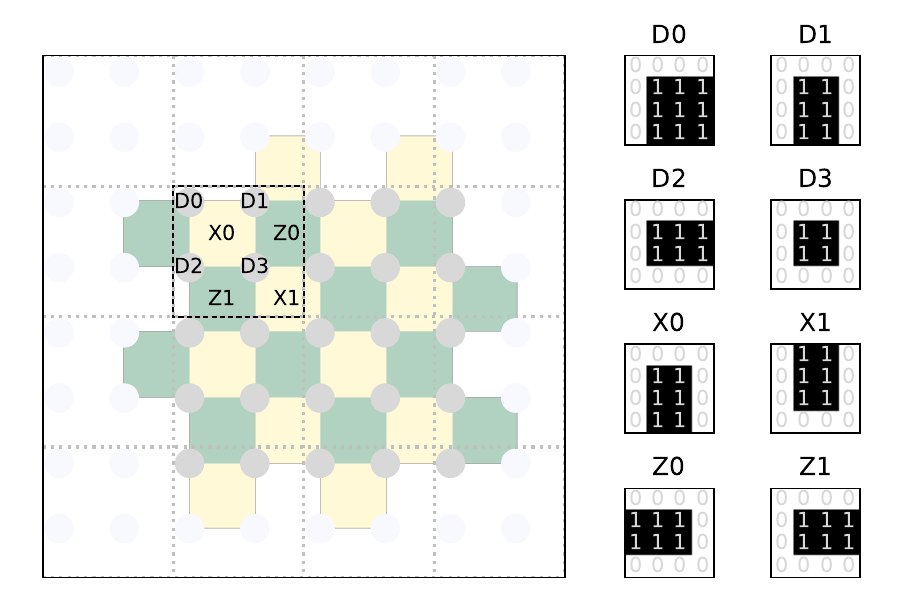}
    \caption{A distance-5 rotated surface code represented using an array of unit cells (offset for clarity). Each unit cell may contain up to four data qubits (grey), two $X$ stabiliser generators (yellow), and two $Z$ stabiliser generators (green). Code boundaries are well defined by specifying the presence of qubits by eight bits per unit cell. Unused data qubits are shown in light grey.}
    \label{fig:1}
\end{figure}

A convenient property of rotated surface codes, and more generally surface codes defined on regular lattices~\cite{PhysRevA.86.020303}, is that data qubits and ancilla qubits can be arranged on a regular 2D lattice. Figure~\ref{fig:1} (a) shows one particular choice of a unit cell for rotated surface codes. Each unit cell contains space for up to four data qubits and up to four stabiliser generators. We note that a smaller primitive cell is possible, containing two data qubits and two stabiliser generators, but requires a large number of empty cells when describing a rotated surface code in a rectangular array. The presence of data qubits and stabiliser generators can be specified by assigning eight bits (presence bits) to each cell of the system, shown for the distance 5 case in Figure~\ref{fig:1} (b). To represent a distance $d$ rotated surface code requires a $n_{r} \times n_{c}$ array of cells, where $n_r=n_c= \lfloor (d+3)/2 \rfloor$.

It would be beneficial to also efficiently represent the state of surface codes which have experienced Pauli noise. Two sets of eight additional bits can be used to keep track of whether each qubits has $X$, $Y$, or $Z$ errors present. This is done by decomposing errors into $X$ components and $Z$ components, ignoring global phase, so that $I$, $X$, $Y$, $Z$ correspond to the bits $00$, $01$, $11$, and $10$, respectively. Continuing with a unit-cell approach, different points in time can be represented by adding an additional dimension to the lattice. In order to maintain a simple meaning for presence bits, $X$ error bits, and $Z$ error bits, the lattice will only represent points in time before the first CNOT of each syndrome measurement cycle. This is possible since Pauli errors can be propagated through CNOT gates using well known circuit identities~\cite{Knill_1998}. As errors are now possible on ancilla bits, 4 additional bits per unit-cell are required to specify whether errors equivalent to measurement errors are found on ancilla qubits. With this description, the decoder-relevant details of the history of a surface code logical qubit suffering circuit noise can be represented in a form where only equivalent errors just before syndrome extraction begins are specified. This motivates an efficient method of parallel simulation which we present in the next section.

\subsubsection{Parallel Simulation}\label{subsec:ch5_parallel_sim}
\begin{figure*}[htp!]
    \centering
    \includegraphics[width=\textwidth]{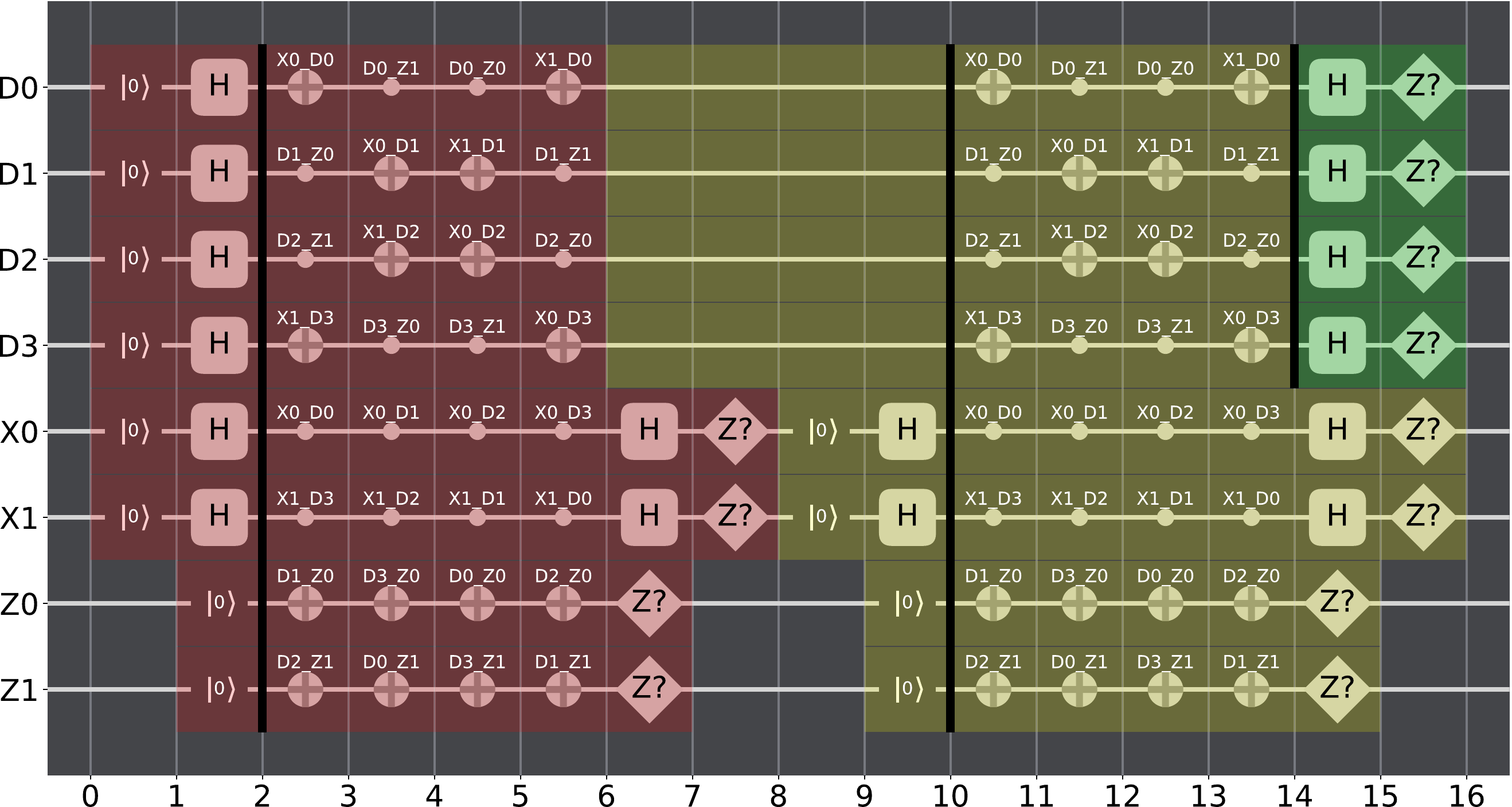}
    \caption{Surface code unit cell syndrome measurement with highlighted error detecting regions and reference time steps as black bars. Gates act between qubits and their nearest neighbours, which sometimes are in adjacent cells. At boundaries, gates scheduled on absent qubits are not applied. The red region highlights a data qubit initialisation round. The yellow region highlights a data qubit idling round. The green region highlights data qubit measurement.}
    \label{fig:ch5_parallel_sim}
\end{figure*}
Surface code simulation is usually performed by simulating syndrome extraction either with stabiliser simulation, or by propagating errors as they occur to find flipped measurement results. The suggestion for representation of logical qubit history described in the previous section motivates a parallel method of simulation. The technique amounts to propagating errors to reference time steps for circuits such as those shown in Figure~\ref{fig:ch5_parallel_sim}. However, as each individual measurement result still depends on the cumulative effect of the light cone of prior events, general measurement outcomes may be unable to be sampled independently. This is rarely a concern for decoding in \gls{qec}, however, where syndrome change events are usually sufficient and are sensitive to Pauli noise of only a finite set of time-steps between two consecutive syndrome measurements. Here we will explain how the data of circuit noise errors and syndrome changes can be found by considering the contribution from different classes of errors separately. These are first described with a separation into ``class 0'', ``class 1'', and ``class 2'' errors \cite{fowler_review_PhysRevA.86.032324}, which correspond to different sets of gate errors, before a decomposition in terms of spacial and temporal correlated syndrome bit flips is instead used.

Class 0 errors describe errors that occur during data qubit idling. The four time-steps of idling errors each data qubit experiences can be combined into a single instance of depolarising noise with parameter ${p_4'}$ using the equation
\begin{equation}\label{eq:depol_noise}
    {p'_n} = 1-(1-p')^n,
\end{equation}
where $p'$ parametrises the intensity of the depolarising noise experienced each time-step individually, setting $n$ to 4. Parametrisation in terms of depolarisation parameters, $p'$, rather than error rates $p=3p'/4$, allows this expression to be easily interpreted as the probability that at least one depolarisation event occurs. Thus, the contribution to error bits and syndrome bits from class 0 errors can be sampled by independently sampling one instance of depolarising noise per qubit, per time-step. One may also consider the additional idling errors of data qubits near boundaries. Under circuit noise models parameterised by depolarisation parameters, these additional errors can be handled by a masked application of class 2 errors, discussed later.

Class 1 errors describe errors associated with ancilla qubit initialisation and measurement. These errors can be considered by simply applying bit-flips to syndrome bits directly. $Z$ stabilisers experience only one instance of reset noise and measurement noise, so experience associated syndrome bit flips with probability ${p'_2}/2$. As $X$ stabilisers also have two additional Hadamard gates, they instead experience syndrome bit flips with  probability ${p'_4}/2$.

Class 2 errors describe those associated with CNOT gates. In circuit noise error models, these are associated with two-qubit depolarising noise after each two-qubit gate. However, as the set of two qubit Pauli gates is invariant under propagation through CNOT gates, class 2 errors can be equally well described by two-qubit depolarising noise just before each CNOT. Pauli errors propagating through more than one CNOT can become higher weight Pauli errors, which demands careful tracking. While some qubits experience CNOT gates, other qubits may have idle time steps when at code boundaries. In place of single qubit depolarising noise, two qubit noise can still be applied on these qubits, where one qubit of the noise channel is absent. Identical error statistics are achieved under uniform noise models if error channels are parametrised by depolarisation parameters $p'$, but not when error channels are parametrised by error rates $p$.

Our particular implementation of parallel simulation decomposes errors into primitive space-like and time-like errors. Each unit cell has up to eight possible space-like errors corresponding to the two $X$ and $Z$ components of errors applied to each of the four data qubit during idle time-steps. Space-like errors flip adjacent anti-commuting stabiliser operator measurements within the same measurement cycle. Space-like errors are also associated with data qubit preparation and measurement errors, as they also flip adjacent stabiliser operators measured on the first syndrome measurement cycle and final set of inferred values from data qubit measurement, respectively. Each unit cell has up to four possible time-like errors associated with erroneous ancilla qubit preparation or measurement. To incorporate the class 2 CNOT errors in this description, we take advantage of the ability to propagate them backwards to find equivalent errors just before the first CNOT of the syndrome measurement cycle. At that point, equivalent errors present on data qubits can be associated with space-like errors and equivalent errors on ancilla qubits can be associated with time-like errors. Care must be taken that global phases are not included as time-like errors, such as when a $Z$ occurs on a freshly initialised $Z$ operator ancilla. Altogether, 12 bits per unit cell are used to specify the equivalent set of new errors that can be thought to have occurred during a particular syndrome measurement cycle.

\begin{figure}[tbp]
    \centering
    \includegraphics[width=\linewidth]{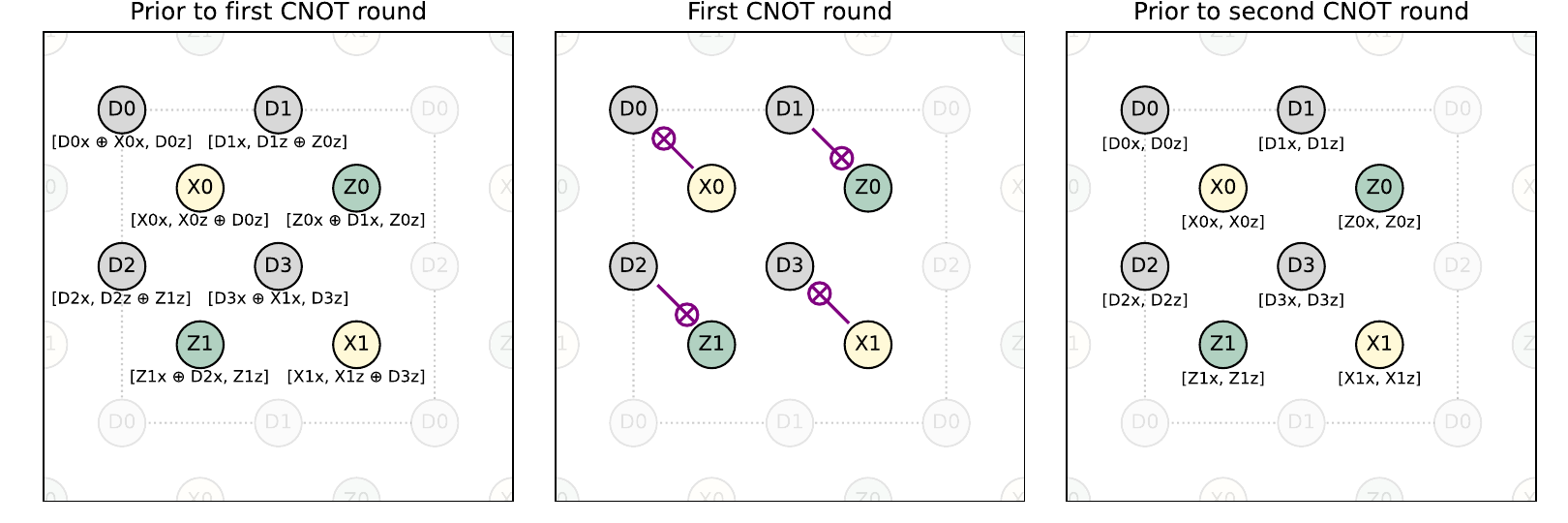}
    \caption{Surface code error bits propagating backwards from before the second CNOT round to before the first CNOT round. Error bits are written in row vectors of form $[E_X, E_Z]$, where $E_X$ and $E_Z$ are $X$ and $Z$ error bits, respectively.}
    \label{fig:ch5_bit_prop}
\end{figure}

The task remains to sample and accumulate all errors to form a complete simulation. Firstly, four random floats are sampled per unit cell to find the contribution from simultaneous data qubit idling time-steps. Error rates are modified to include contributions from data qubit initialisation and measurement errors during the initial and final time-steps, respectively. These space-like errors are stored in two $X$ and $Z$ space-time error bits associated with the respective qubits. Next, four random floats are sampled per unit cell to find the contribution from ancilla qubit preparation and measurement errors. Differences in $X$ and $Z$ ancilla effective error rates are taken into account. These time-like errors are stored in the four $X$ and $Z$ time-like error bits per cell. Lastly, 16 random floats are sampled per unit cell which corresponds to whether errors occurred during each of the four time-steps of four CNOT gates per unit cell. After sampling according to the error rate, the 16 resultant bits are repeated 4 times to form a mask of 64 bits per unit cell. When this mask is applied to a new sampling of 64 uniformly random bits, the resultant set of 64 bits correspond to the correlated $X$ and $Z$ errors associated with parallel sampling of circuit noise interpreted to occur just before each CNOT gate. The bits associated with missing qubits can be set to zero with the use of presence bit information. The bits which correspond to qubits near edges, which do not have a partner to interact with during some CNOT time-steps, are used as idling error bits, which is a convenient property of noise models parametrised by uniform depolarisation parameters $p'$. The final step involves propagating all error bits to their equivalent values just before the first CNOT. This was done by performing vectorised boolean logic operations on error bits of future time-steps. Alternatively, parallel error propagation may be performed with the use of 2D convolutional kernels. An example of a bit propagation is shown in Figure \ref{fig:ch5_bit_prop}. Making use of the periodicity of surface codes, multiple qubits can have their error states upgraded simultaneously, based on simple boolean functions of their prior (or in this case future) states.

Finally, the associated syndrome change events must be calculated. This is calculated in parallel by summing up the error bits associated with each syndrome change site (modulo 2). The space-time errors and syndrome changes can be plotted together in 3D as shown in Figure~\ref{fig:decoder_demands} (b).

Before proceeding to the application of this simulation technique to decoder training, the connection to decoder graphs and \glspl{dem} should be noted. The parallel simulation described here can be essentially thought of as sampling edges from a decoder hyper-graph, similar to what was recently described by Chamberland \etaldot~\cite{Chamberland_2023}, or sampling flip events from a \gls{dem}~\cite{Gidney2021stimfaststabilizer}. The propagation of errors to just before the first CNOT of syndrome extraction is the key additional result, which allows decompositions of equivalent space-like and time-like errors to be ascribed to outcomes of sampling circuit noise. This defines an explicit mapping from circuit noise to correlated phenomenological noise.

\subsubsection{Data Augmentation}\label{subsec:ch5_simplifiers}
\begin{figure*}[htp!]
    \centering
    \includegraphics[width=0.45\textwidth]{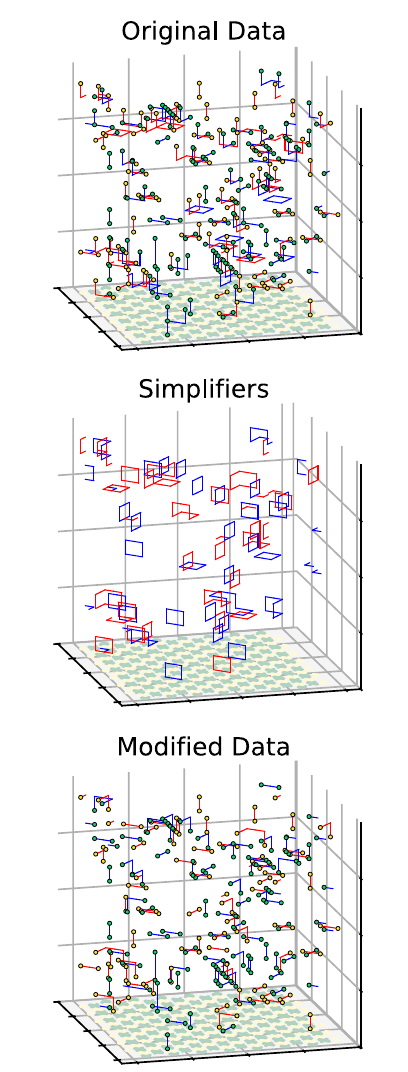}
    \caption{The application of simplifier operators on rotated surface code data. The subfigure titled ``Original Data'' shows errors and syndrome bit changes of a distance 15 code with $p'=0.005$ noise. The subfigure titled ``Simplifiers'' shows error chain loops identified to reduce total error count or break symmetry. The subfigure titled ``Modified Data'' shows the resultant set of edges after applying simplifiers.}
    \label{fig:ch5_simplifications}
\end{figure*}
Surface code space-time edges can be modified while preserving the effective errors experienced. For example, when stabiliser operators are applied to surface code code-space states during a particular instance of time, the resultant state is identical to the initial state. If instead stabiliser operators were applied to a surface code code-space state with Pauli noise, a global phase of $-1$ may be applied, which can usually be ignored. From the perspective of space-time edges, this corresponds to applying a bit-wise XOR operation between the original space-time edges and a set of space-time edges containing a loop parallel to the $X-Y$ plane (or starting and ending at the same boundary). These space-time edges correspond to a trivial syndrome and do not apply a non-trivial logical operation. We can lift the restriction to the $X-Y$ plane to instead consider all loops of space-time errors, which again result in a trivial syndrome and no net nontrivial logical operators. We will refer to these space-time edges as ``space-time simplifiers'' or just ``simplifiers'' for short. Space-time simplifiers can be interpreted as modifications to errors which may have occurred but never have any measurable effects. A set of space-time edges forming a basis for space-time simplifiers can be constructed by considering the application of each different stabiliser operator at all points in time (a total of 4 per unit cell), and also applying each different pair of repeated errors together with adjacent time-like errors on non-commuting stabiliser qubits (a total of 8 per unit cell). The set of simplifiers form a group, where each element is its own inverse, and composition can be defined by addition mod 2 when space-time errors are represented with bits. Associativity and commutativity follow from the definition of the composition operation. 

Simplifiers can be applied to reduce the number of nontrivial edges in a space-time edge representation. This can be done by checking whether primitive simplifiers have more than half of their edges activated, and applying rule-based symmetry breaking similar to what was described in previous work \cite{Gicev2023scalablefast}. Finding optimal solutions can be expected to be prohibited, in most cases, by in general needing to check an exponential number of distinct primitive simplifier combinations. An example of the effect of simplifier operations on sampled circuit noise close to threshold is shown in Figure~\ref{fig:ch5_simplifications}. From this example, we can see that surface code error data close to threshold frequently contains simplifiers in the form of full space-like stabiliser loops, and features significant variation in how error chains connecting two changed syndrome bits are sampled. Both of these effects are expected to reduce \gls{ann} decoder performance.

\clearpage
\subsection{Decoder Performance}\label{subsec:ch5_ann_approaches}
\subsubsection{PyMatching Decoders}
\begin{figure}[b]
    \centering
    \includegraphics[width=\textwidth]{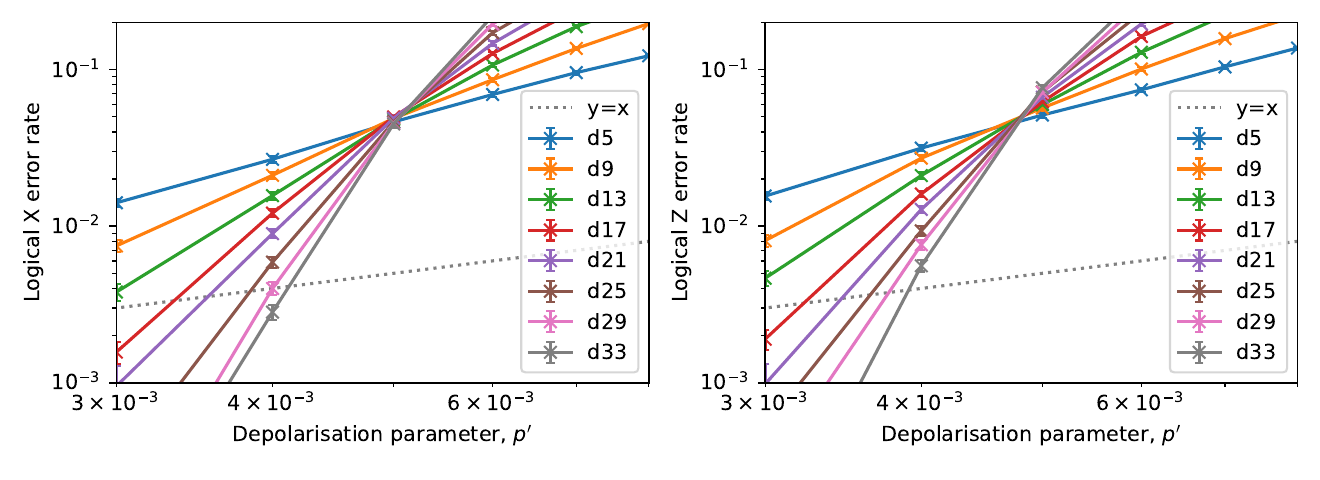}
    \caption{Performance of \gls{pcm} PyMatching for rotated surface codes suffering uniform depolarisation parameter, $p'$, noise.}
    \label{fig:base_mwpm}
\end{figure}

PyMatching is a Python implementation of \gls{mwpm} commonly used in \gls{qec} research due its relative ease of use and compatibility with the optimised stabiliser simulator Stim~\cite{Gidney2021stimfaststabilizer, Higgott2021arxiv}. Here we will discuss two decoders constructed using PyMatching which will later be used as mop-up decoders for \gls{ann} models. The first, referred to as \gls{pcm} PyMatching, is constructed using a matrix defined by the code stabilisers. This then can be used to define a rectilinear matching problem, where diagonal hook errors are not present, effectively ignoring many effects associated with circuit noise such as hook errors and relative error likelihoods. The second, referred to as \gls{dem} PyMatching, is constructed utilising a \gls{dem} derived from the noisy Stim circuits each of memory experiment. \glspl{dem} refer to a set of probabilities assigned to combinations of detection events and logical operator changes caused by independent error events. This allows the construction of decoding graphs which include diagonal hook errors, although still omit some correlations caused when errors cause more than 2 detection events. Results for $n_c=d$ rounds of decoding for distance $d=5$ to $d=33$ codes using \gls{pcm} PyMatching are shown in Figure~\ref{fig:base_mwpm}. A threshold at approximately $p'=5\times10^{-3}$ is attained, with logical noise biased towards $Z$ errors. The bias towards $Z$ errors can be understood to arise due to $X$ stabiliser measurement circuits containing two more instances of depolarising noise than $Z$ stabiliser measurement circuits.

\begin{figure}[ht]
    \centering
    \includegraphics[width=\textwidth]{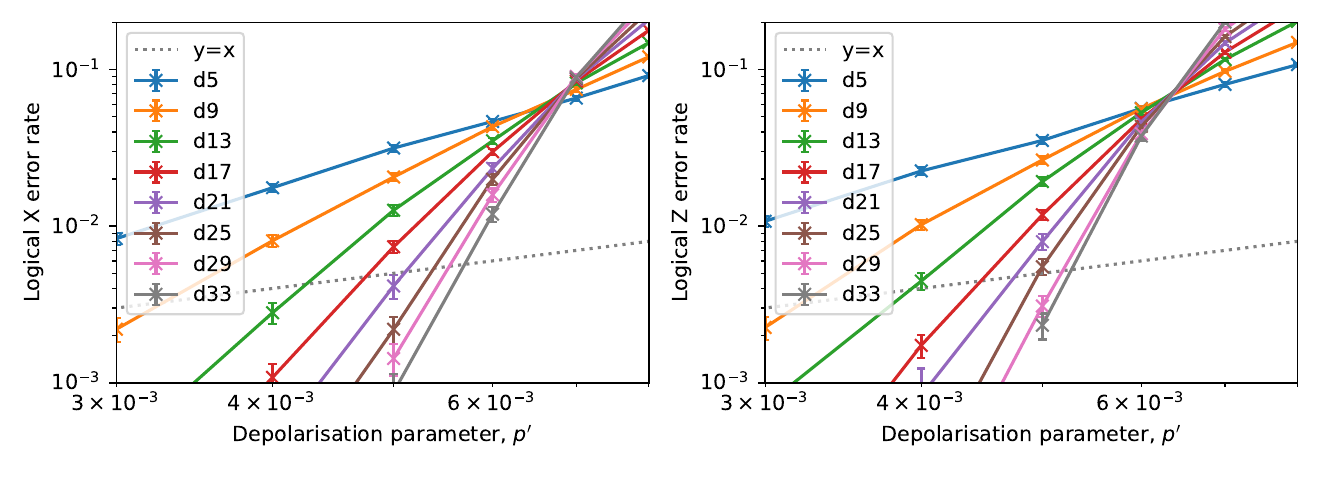}
    \caption{Performance of \gls{dem} PyMatching for rotated surface codes suffering uniform depolarisation parameter, $p'$, noise.}
    \label{fig:dem_mwpm}
\end{figure}

Results for $n_c=d$ rounds of decoding for distance $d=5$ to $d=33$ codes using \gls{dem} PyMatching are shown in Figure~\ref{fig:dem_mwpm}. The results show an increased threshold exceeding $p'=6.5\times10^{-3}$, and a corresponding substantial decrease in \glspl{ler} below threshold compared to \gls{pcm} PyMatching. Noise remains biased towards $Z$ errors. Although the optimisation of edge weights and inclusion of hook error edges is generally considered to not demand substantial additional computational overhead, the use of these optimisations does rely on assumptions about the noise present, which may not be justified on a real quantum device, although \gls{ann}-based training can also suffer the same pitfall. Additionally, the performance of \gls{pcm} PyMatching is still of relevance as a point of comparison when it is used as a global decoder for \gls{ann}-based decoding.

\begin{figure}[p]
    \centering
    \includegraphics[ width=\textwidth]{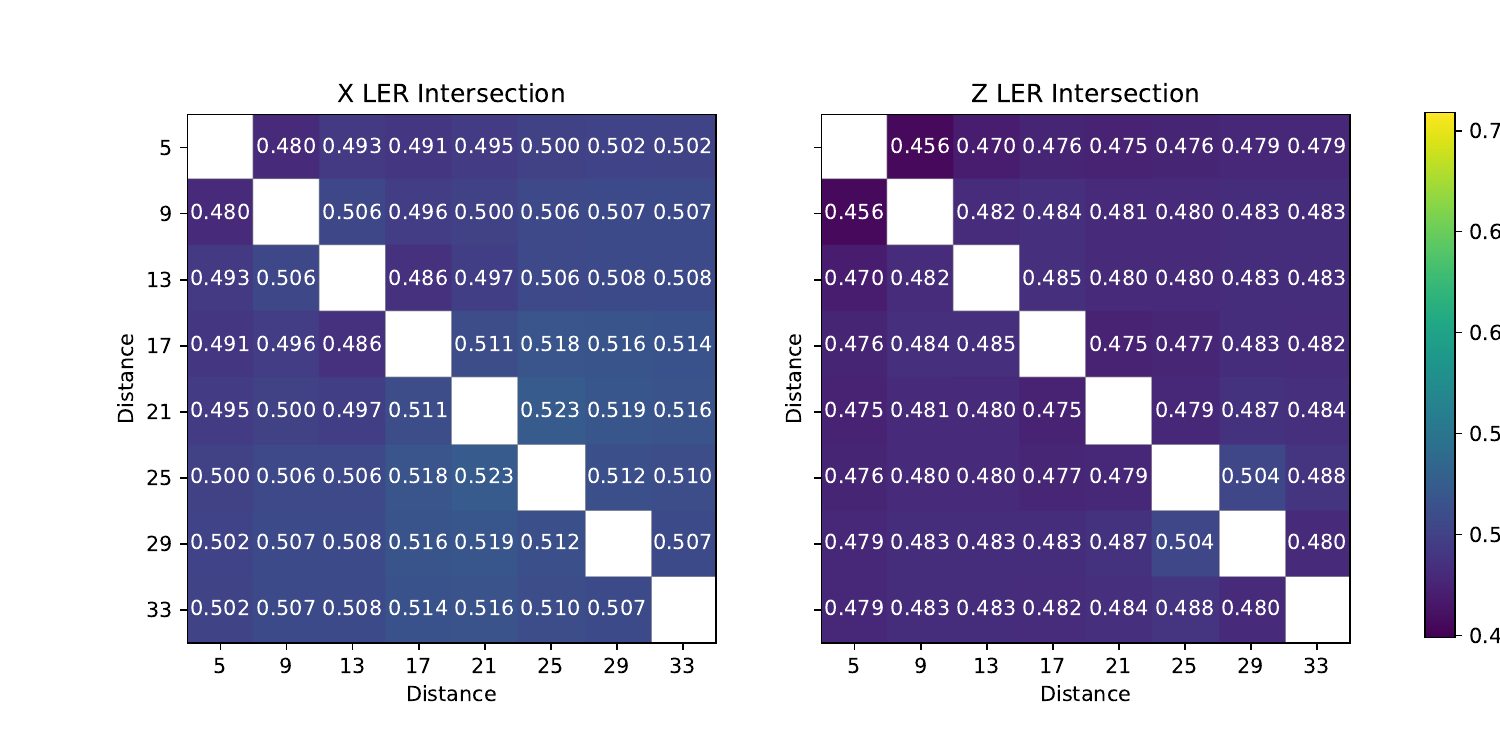}
    \caption{Matrices showing intersections between \gls{ler} curves for rotated surface codes of distances between 5 and 33 when decoded with \gls{pcm} PyMatching. Text in each cell shows the depolarisation parameter value rounded to three decimal places based on extrapolation of local error rates calculated at depolarisation parameters at the nearest two tenth of a percent.}
    \label{fig:base-intersection}
\end{figure}

\begin{figure}[p]
    \centering
    \includegraphics[ width=\textwidth]{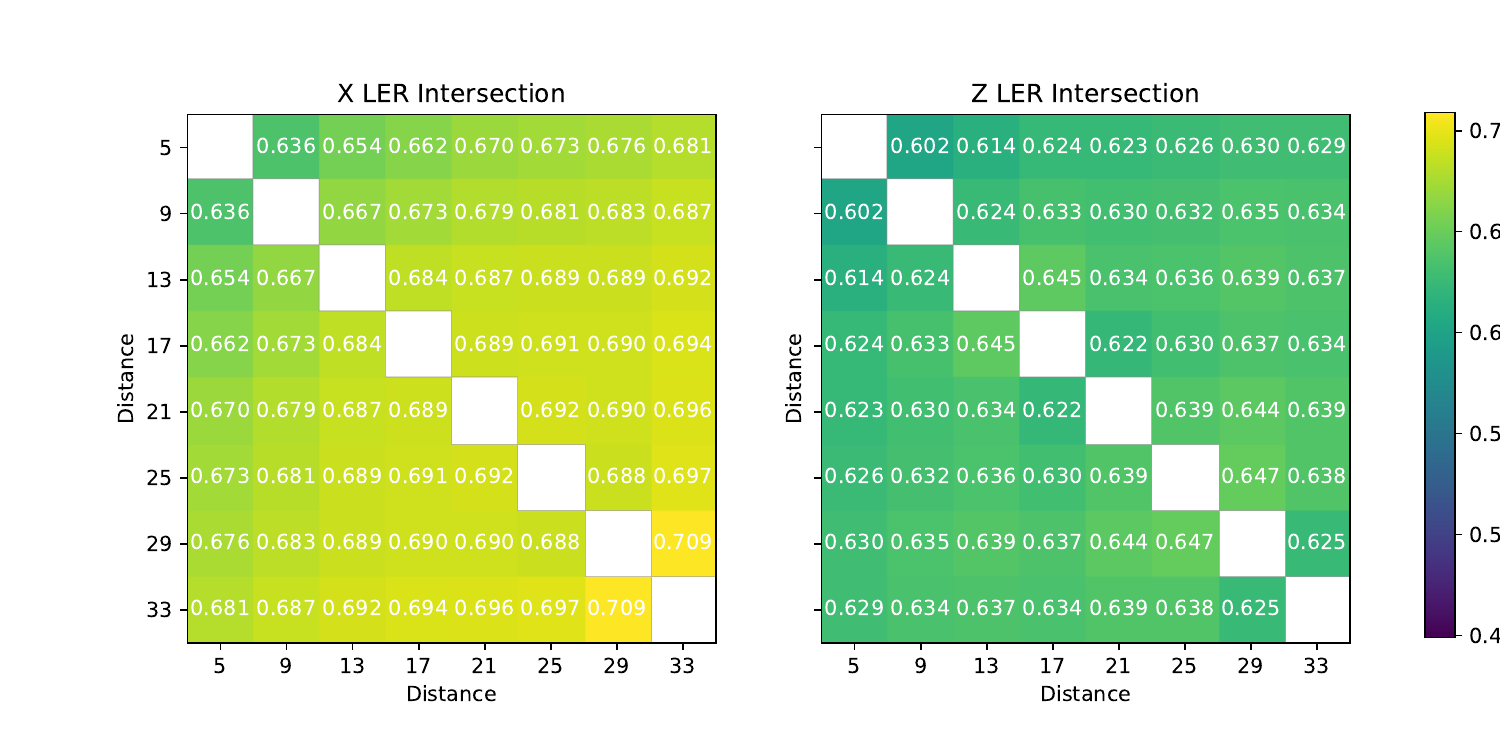}
    \caption{Matrices showing intersections between \gls{ler} curves for rotated surface codes of distances between 5 and 33 when decoded with \gls{dem} PyMatching. Text in each cell shows the depolarisation parameter value rounded to three decimal places based on extrapolation of local error rates calculated at depolarisation parameters at the nearest two tenth of a percent.}
    \label{fig:dem-intersection}
\end{figure}

An interesting property of circuit noise threshold curves for PyMatching decoders is that intersections between different code distances appear close to the same depolarisation parameter value. A variation in terms of code distance is to be expected, as lower codes are dominated by boundary effects which are gradually overtaken by bulk effects for larger distances. To investigate this phenomenon further, we calculated exact intersection points assuming power-law extrapolation is valid near the intersection points. This corresponds to linear extrapolation on the log-log plot. Results are show in Figures~\ref{fig:base-intersection} and~\ref{fig:dem-intersection} for \gls{pcm} and \gls{dem} PyMatching, respectively. From the two figures, we note that intersection points indeed tend to increase as code distances increase, suggesting that threshold error rates are likely to be underestimated. Additionally, it can be observed that depolarisation parameters of intersections of $Z$ \gls{ler} curves are consistently lower than those of the corresponding $X$ \gls{ler} curves. This can be understood to follow from the asymmetry inherent in the syndrome measurement circuits when Hadamard change of basis operations are used~\cite{ashleypra2014}.

 \clearpage
\subsubsection{Multi-label Classifier Decoder}\label{sec_ch5:multilabel}
In order to take advantage of translational symmetry inherent in surface code decoding, the neural network architecture we considered corresponds to repeated three-dimensional convolution layers. Activation functions were set to ReLu activations, with sigmoid activation functions on the last layer. This ensures that the final layer can be interpreted as a probability distribution of error correction bits, conditioned on the syndrome bits and boundary information given as input. Three hidden layers were used, each featuring 128 kernels and kernel sizes of $3\times3\times3$. Training data was generated by simulating surface code memory experiments via vectorised unit-cell based simulation with uniform depolarisation parameter noise channels. Approximately $500,000$ training instances were generated, corresponding to distance $33$ rotated surface codes suffering circuit noise with depolarisation parameter values from $0.001$ to $0.007$.

\begin{figure}[htbp]
    \centering
    \includegraphics[ width=\textwidth]{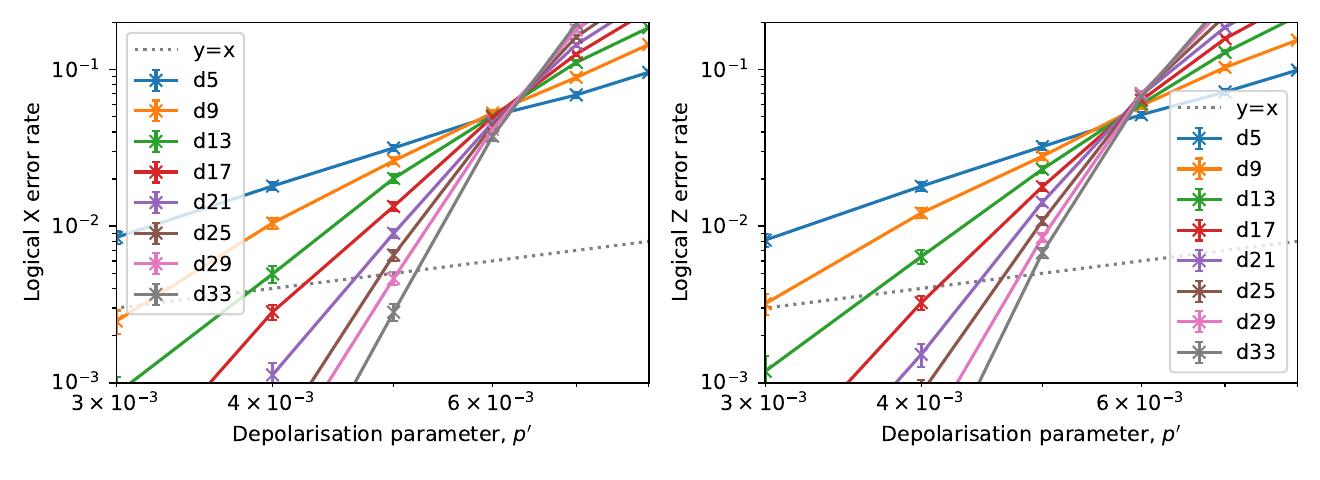}
    \caption{Performance multi-label classifier \gls{ann} decoding, followed by \gls{pcm} PyMatching for rotated surface codes suffering uniform depolarisation parameter, $p'$, noise.}
    \label{fig:ann_base_mwpm}
\end{figure}

\begin{figure}[htbp]
    \centering
    \includegraphics[ width=\textwidth]{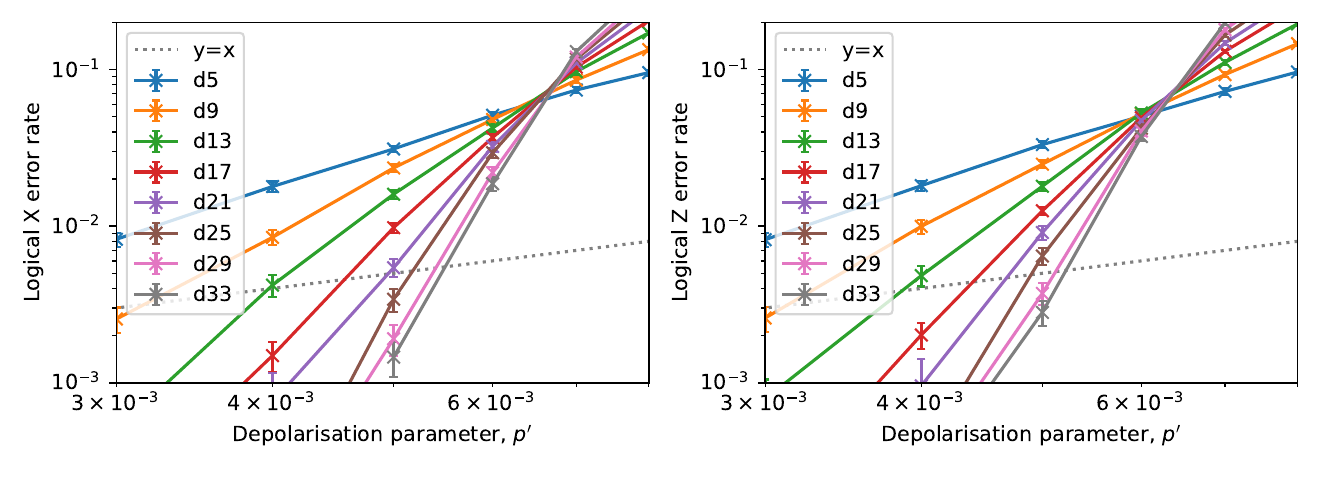}
    \caption{Performance multi-label classifier \gls{ann} decoding, followed by \gls{dem} PyMatching for rotated surface codes suffering uniform depolarisation parameter, $p'$, noise.}
    \label{fig:ann_dem_mwpm}
\end{figure}

Testing was performed by sampling syndrome strings and logical operator changes in Stim~\cite{Gidney2021stimfaststabilizer}. \gls{ann} corrections were obtained by providing boundary information and syndrome information as input and receiving corrections in the form of space-like and time-like errors as output. The syndrome bits associated with the output corrections were then used to calculate changes to the original syndrome using local parity sums. The resultant residual syndrome differs from the original syndrome by some syndrome bits being resolved (matched with other bits or to a boundary) and some instead displaced. Residual syndrome bits are decoded with an auxiliary global decoder, in this case PyMatching, which outputs whether a logical operator is expected to be crossed after matching is performed. Results are shown in Figures \ref{fig:ann_base_mwpm} and \ref{fig:ann_dem_mwpm} for \gls{pcm} and \gls{dem} PyMatching mop-up, respectively. We find that the use of \gls{ann} decoding offers \gls{ler} and threshold improvements for \gls{pcm} PyMatching. In contrast, performance for \gls{dem} PyMatching appears to generally not show improvement compared to using the global decoder alone.


\subsubsection{Simplified Multi-label Classifier Decoder}
\begin{figure}[htbp]
    \centering
    \includegraphics[width=0.45\textwidth]{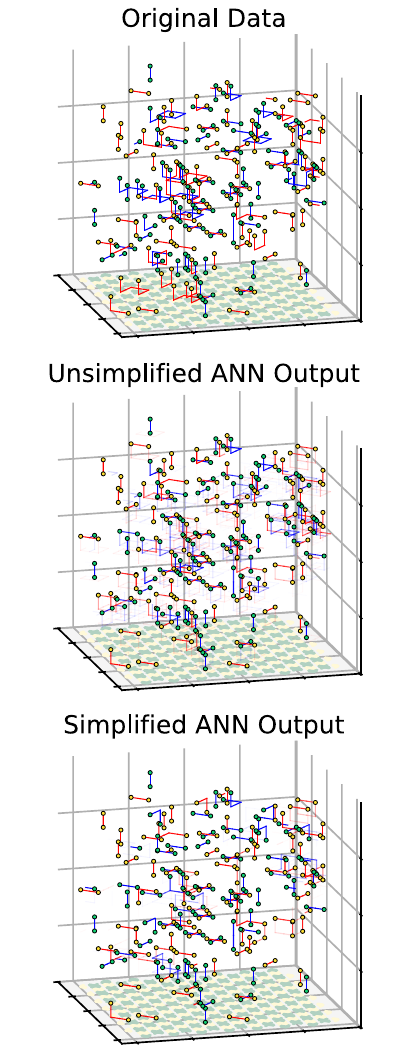}
    \caption{Output predictions of \glspl{ann} when trained on unsimplified data compared to simplified data. The original data corrections to circuit noise with depolarisation parameter $p=0.005$.}
    \label{fig:simplified_vs_unsimplfied_prediction}
\end{figure}

An identical neural network structure was trained on data having gone through the error simplification process described in Section~\ref{subsec:ch5_simplifiers}. The use of error simplification reduces the negative impacts associated with full error loops and variety in error chains. The result is a network which predicts error chains with increased certainty, providing output probabilities away from $0.5$. An example showing the difference between outputs for networks trained with unsimplified data compared to simplified data is shown in Figure~\ref{fig:simplified_vs_unsimplfied_prediction}. The key property is the reduced tendency to split error probabilities across multiple equivalent error chains when symmetry is broken by the error simplification process.

\begin{figure}[t]
    \centering
    \includegraphics[ width=\textwidth]{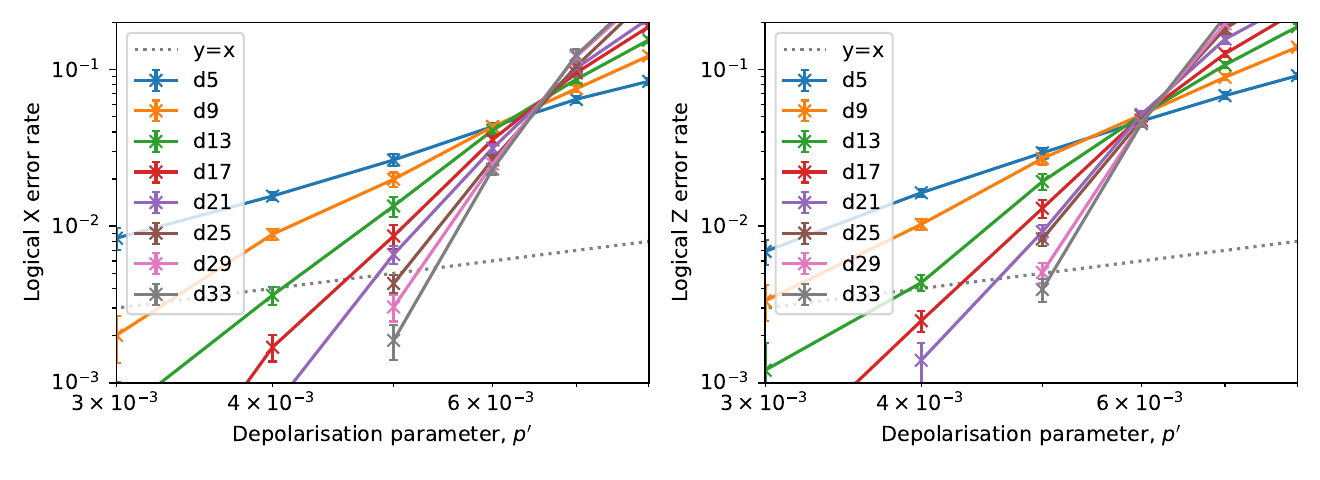}
    \caption{Performance multi-label classifier \gls{ann} (simplified data during training) decoding, followed by \gls{pcm} PyMatching for rotated surface codes suffering uniform depolarisation parameter, $p'$, noise.}
    \label{fig:multilabel_simplified_base}
\end{figure}

\begin{figure}[t]
    \centering
    \includegraphics[ width=\textwidth]{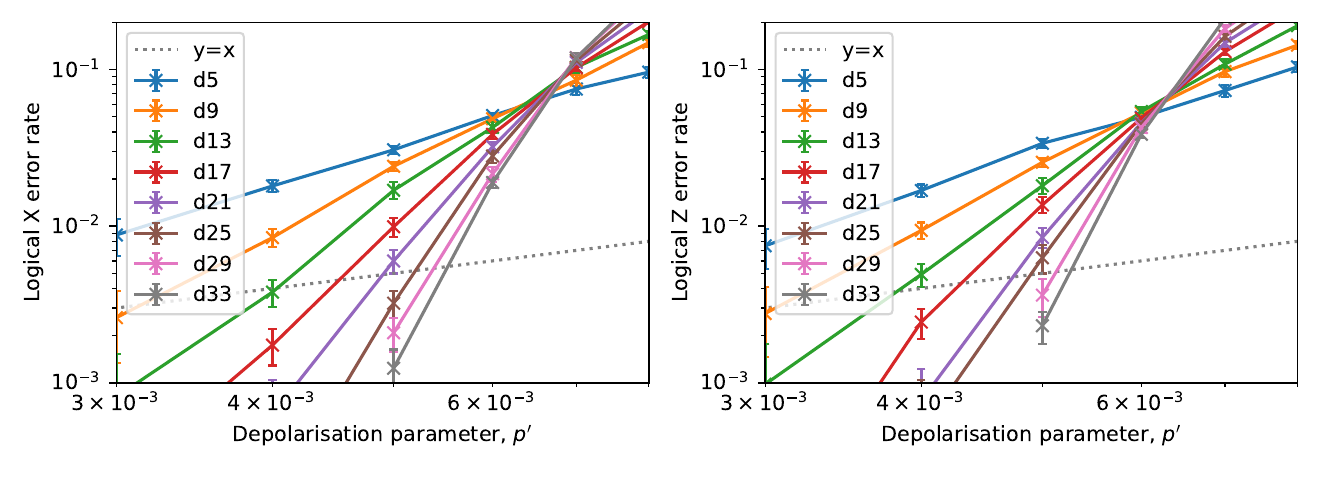}
    \caption{Performance multi-label classifier \gls{ann} (simplified data during training) decoding, followed by \gls{dem} PyMatching for rotated surface codes suffering uniform depolarisation parameter, $p'$, noise.}
    \label{fig:multilabel_simplified_dem}
\end{figure}

Utilising error simplification was observed to improve \glspl{ler} and increase thresholds in previous works \cite{Chamberland_2023}. Figures~\ref{fig:multilabel_simplified_base} and~\ref{fig:multilabel_simplified_dem} show \gls{ler} curves when using \gls{pcm} PyMatching and \gls{dem} PyMatching as a mop-up decoders, respectively. From the figures, it can be seen that there are improvements to \glspl{ler} and thresholds when compared with the related results for unsimplified data. The improvements are, however, less pronounced for the comparison with the \gls{dem} PyMatching decoder. It should be noted that this improved performance comes at the cost of no additional demands for \gls{ann} computation at run time, as the network structures are identical. The simplification process does, however, come at the one time cost of increased computational demands during training, as searching for and applying simplification operations requires comparable operations to vectorised simulation.

 \clearpage
\subsubsection{Conditional Diffusion Decoder}
Beyond a multi-label classification approach, additional \gls{ml} paradigms have been developed which may have relevance to the decoding problem of \gls{qec} codes. One example corresponds to mapping the problem of generating adequate \gls{qec} corrections to generative modelling. Diffusion models have seen recent popularity in generative modelling tasks \cite{NEURIPS2020_4c5bcfec}, but have not yet been investigated in the context of \gls{qec} decoding. We used our training data to train a similarly structured neural network as our multi-label classification models to instead perform the task of diffusive generative modelling. The input layer of the diffusion model was changed to include a source of tentative corrections, initialised as uniform random values between zero and one. Additionally, the syndrome associated with the tentative corrections is calculated by generalising parity calculations using fuzzy logic. This corresponded to repeated application of fuzzy XOR operations, defined as
\begin{equation}
    x\oplus_{\mathrm{fuzzy} }y = x(1-y)+(1-x)y,
\end{equation}
where the inputs $x,y\in[0, 1]$ correspond to tentative correction values. A fuzzy residual syndrome is finally calculated by calculating the bit-wise fuzzy XOR between the original syndrome and the fuzzy syndrome of the tentative corrections. This is given as an additional input to the network to give information on where further modifications to corrections may be needed. Once this training data is prepared, \gls{ann} training can proceed as a supervised learning problem. We note that in contrast to most diffusion models, the model presented here was trained to predict samples of original errors directly, rather than modifications to the tentative corrections. Additionally, no time schedule was used and noise was added to the original errors of the training data by mixing the data with noise scale parameter $p_{mix}\in[0, 1]$. The model was trained on approximately $500,000$ training instances of distance $33$ rotated surface codes suffering circuit noise depolarisation parameter values from $0.001$ to $0.007$. We note that no simplifications to remove loops or break symmetry was applied to the training data.

\begin{figure}[htbp]
    \centering
    \includegraphics[width=\textwidth]{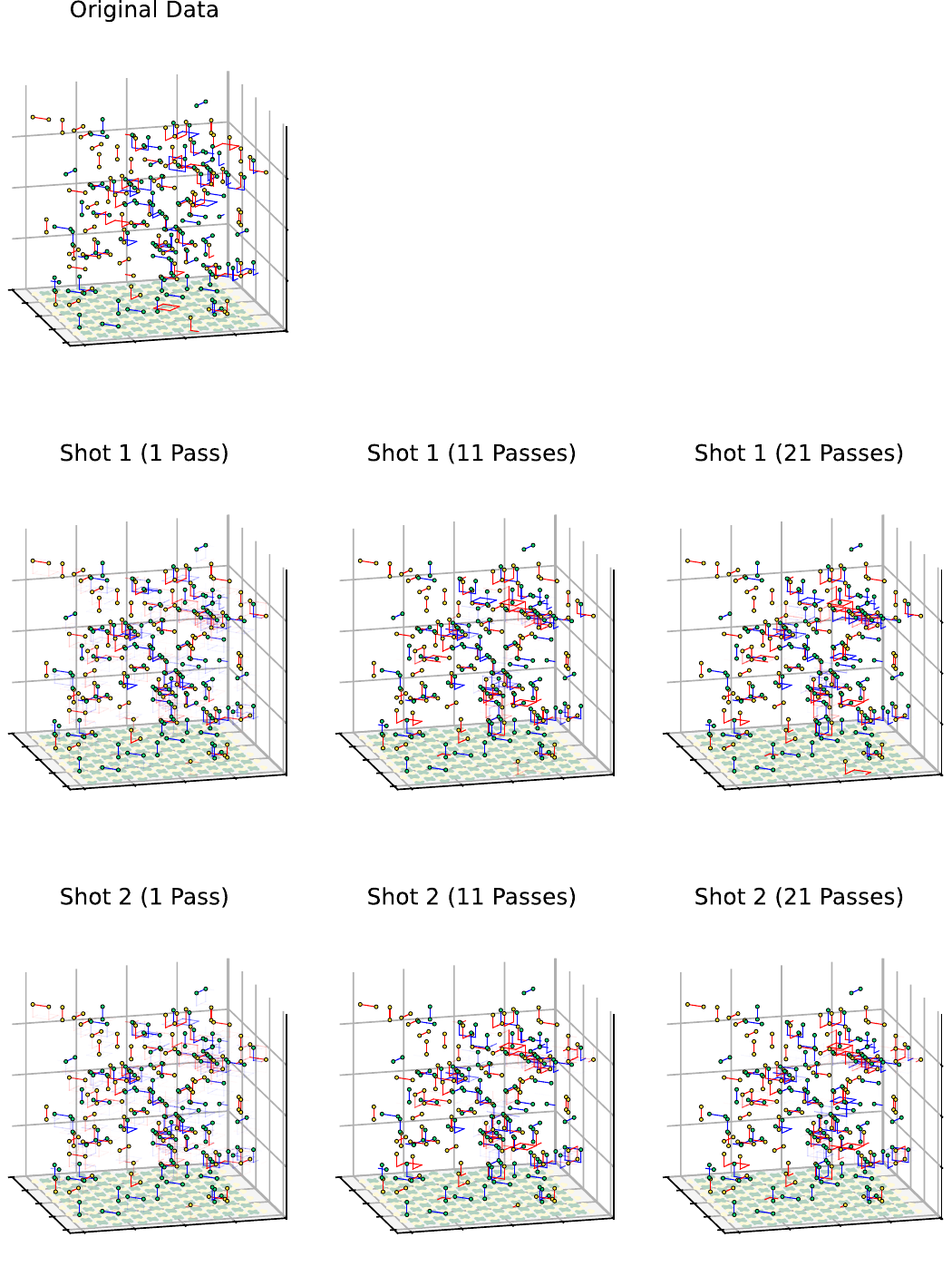}
    \caption{Output predictions a diffusion \gls{ann} after 1, 11 and 21 passes through the model. Two shots are shown, corresponding to different random noise given as input. The original data corrections to circuit noise with depolarisation parameter $p'=0.005$.}
    \label{fig:diffusion-example}
\end{figure}

An interesting characteristic of generative models is that they should be able to produce non-deterministic outputs. This means that, unlike the multi-label classifier decoders which always produce the same set of corrections, repeated sampling of a diffusion model should result is different sets of corrections. Figure \ref{fig:diffusion-example} shows two shots of \gls{ann} corrections performed for the same syndrome, but different random tentative corrections as input. We note that we do observe slight differences in corrections based on input noise given to the network. However, the network still regularly provides the same output, in spite of different noise inputs, in particular near isolated individual errors. We observed that a network trained to perform the diffusion task yields similar uncertainty in output predictions (multiple additional faint edges) as multi-label classification models, but with reduced uncertainty after multiple additional passes. Interestingly, this includes prediction of superfluous corrections in the form of loops. The results show that residual corrections are still sometimes needed from a global decoder.

\begin{figure}[t]
    \centering
    \includegraphics[width=\textwidth]{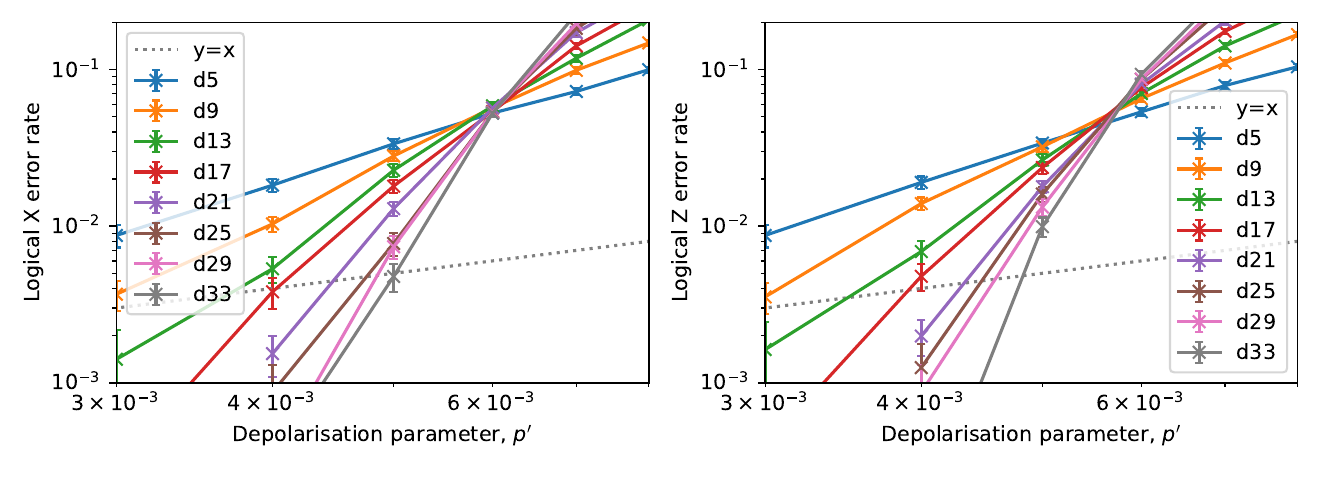}
    \caption{Performance 11-pass diffusion \gls{ann} decoding, followed by \gls{pcm} PyMatching for rotated surface codes suffering uniform depolarisation parameter, $p'$, noise.}
    \label{fig:base-diffusion}
\end{figure}

\begin{figure}[t]
    \centering
    \includegraphics[ width=\textwidth]{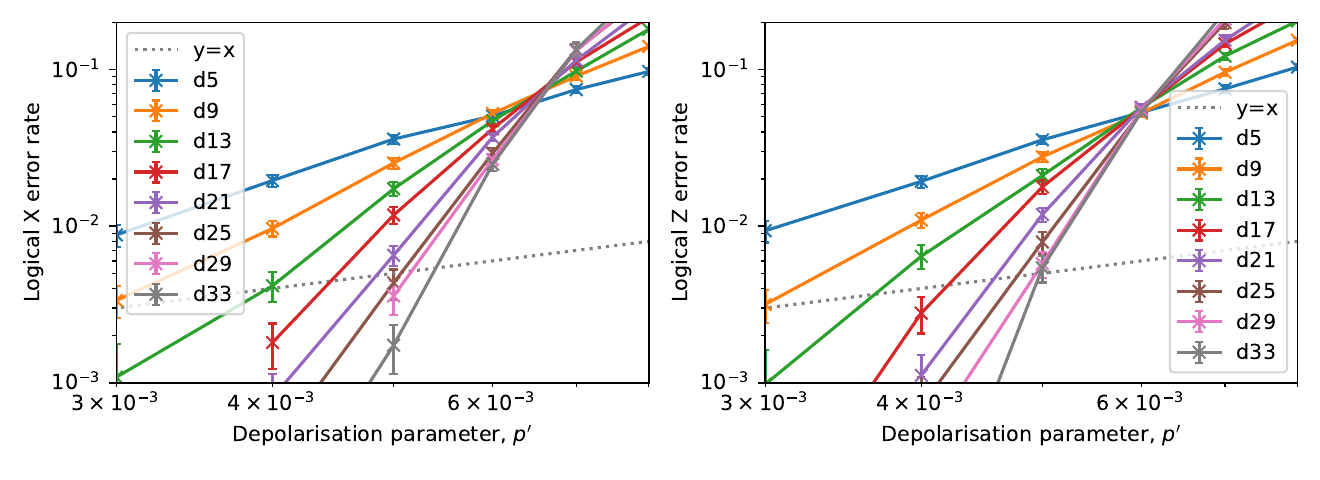}
    \caption{Performance 11-pass diffusion \gls{ann} decoding, followed by \gls{dem} PyMatching for rotated surface codes suffering uniform depolarisation parameter, $p'$, noise.}
    \label{fig:dem-diffusion}
\end{figure}

The \gls{ler} performance of our diffusion model decoder is shown in Figures \ref{fig:base-diffusion} and \ref{fig:dem-diffusion} when performing mop-up decoding with \gls{pcm} PyMatching and \gls{dem} PyMatching, respectively. We find that comparable performance is achieved compared to the models based on multi-label classification.

\begin{figure}[htbp]
    \centering
    \includegraphics[width=\textwidth]{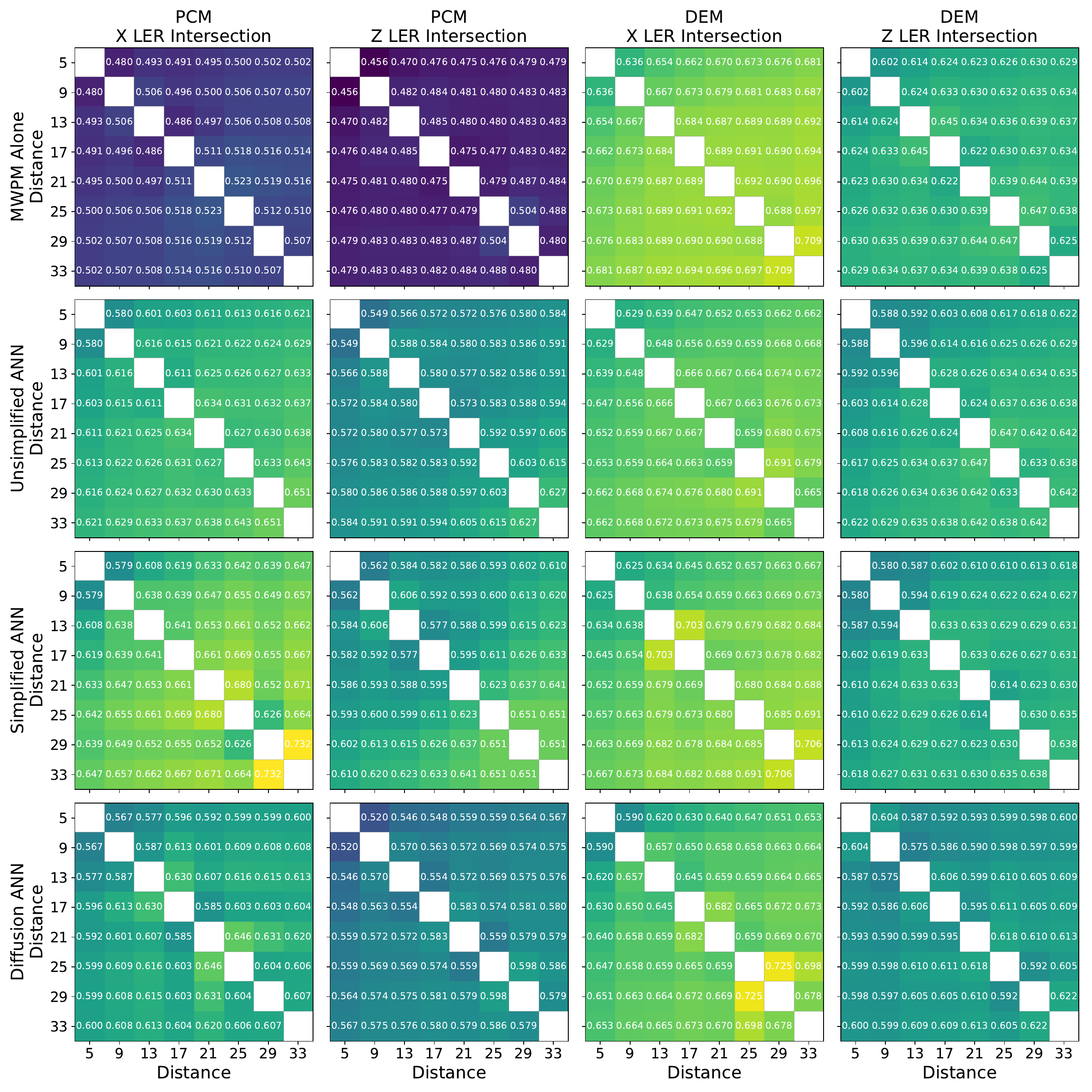}
    \caption{Matrices showing intersections between \gls{ler} curves for rotated surface codes of distances between 5 and 33 when decoded with \gls{pcm} or \gls{dem} PyMatching (\gls{mwpm}) alone and the three \gls{ann} decoders presented in this work. Text in each cell shows the depolarisation parameter value rounded to three decimal places based on extrapolation of local error rates calculated at depolarisation parameters at the nearest two tenth of a percent.}
    \label{fig:intersection_all}
\end{figure}

As a final comparison between all three \gls{ann} decoders presented here, the intersection depolarisation parameters between difference distance codes is shown in Figure \ref{fig:intersection_all}. The performance of the \gls{mwpm} decoders each \gls{ann} model uses for the mop-up step is shown again in the top row for convenience. The results show that \gls{pcm} PyMatching benefits from the use of each of the three \gls{ann} decoders, with slightly better performance on average achieved when using simplified training data. For \gls{dem} PyMatching, the results consistently show modest performance improvements for $X$ \glspl{ler}, with competitive, but mixed, results for $Z$ \glspl{ler}.

 \clearpage
\subsection{Decoding Timing}\label{subsec:ch5_timing}
\begin{figure}[t]
    \centering
    \includegraphics[width=\linewidth]{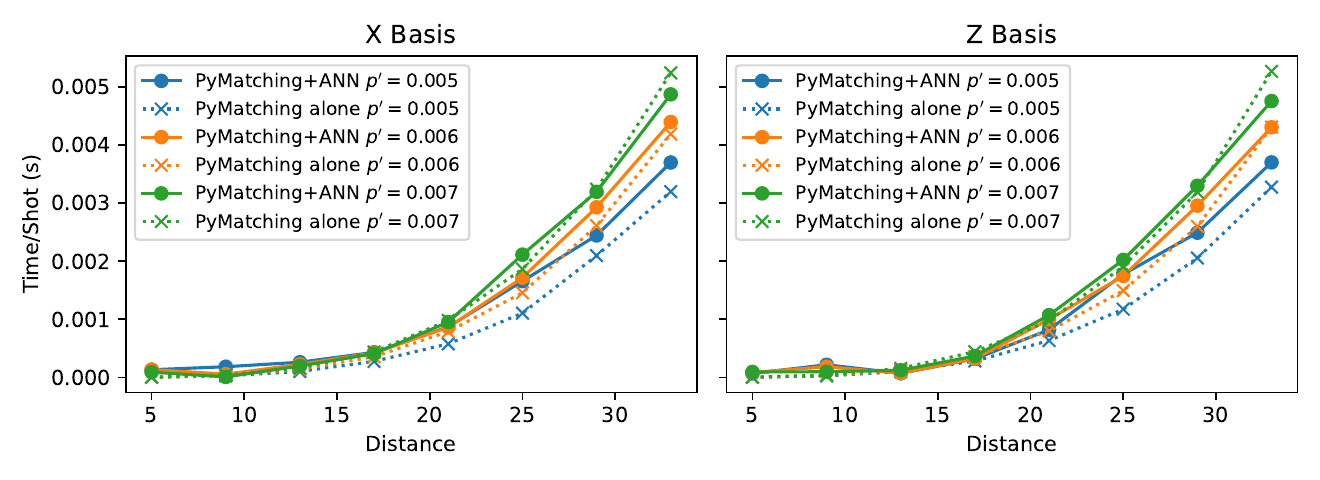}
    \caption{Decoding times for \gls{pcm} PyMatching decoding compared with multi-label \gls{ann} decoding with \gls{pcm} PyMatching for $X$ and $Z$ basis memory experiments with circuit noise depolarisation parameters near threshold, $p'_{\mathrm{th}}\approx0.006$.}
    \label{fig:ch5_timing}
\end{figure}

\begin{figure}[htbp]
    \centering
    \includegraphics[width=\linewidth]{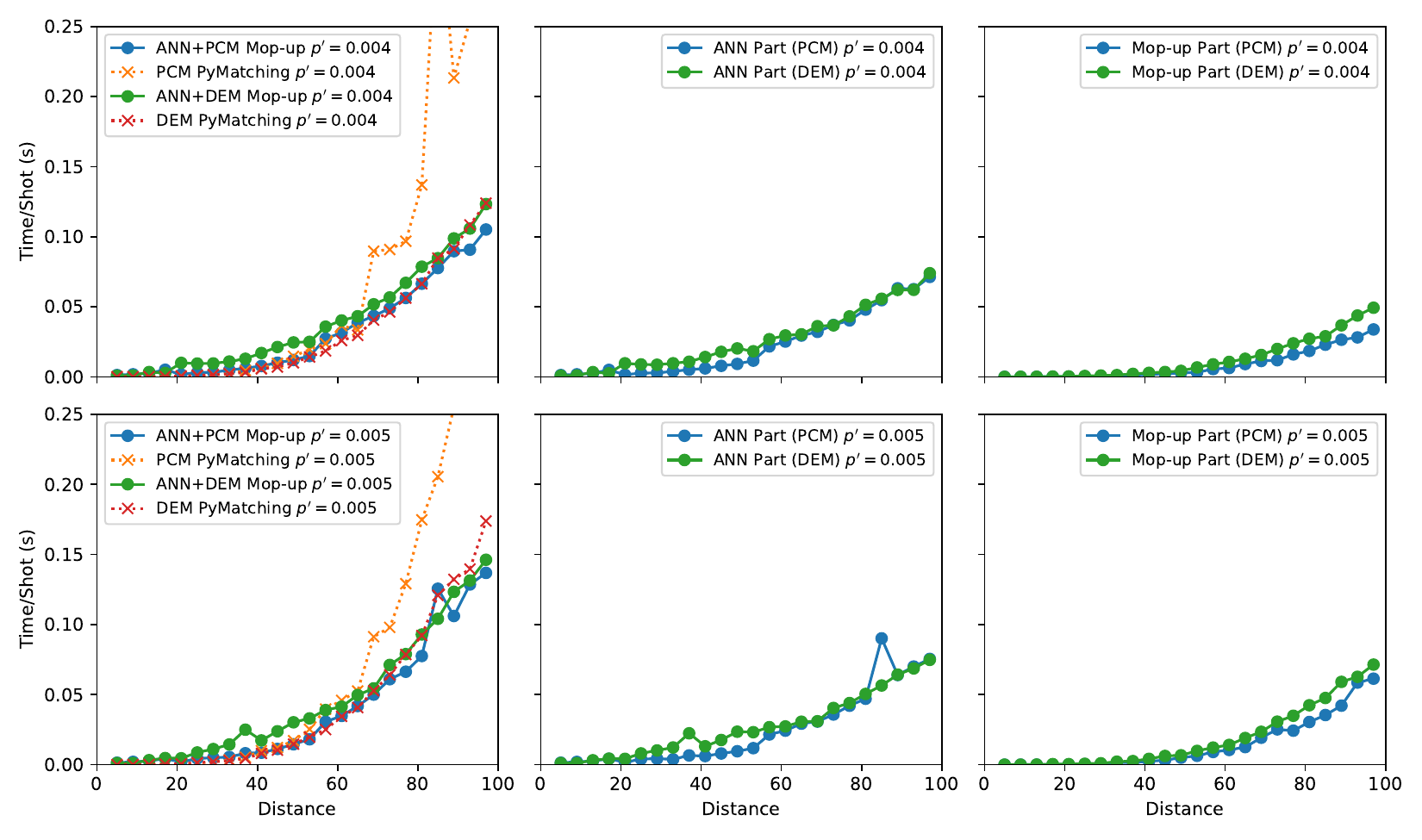}
    \caption{Decoding times for \gls{pcm} PyMatching and \gls{dem} PyMatching decoding compared with multi-label \gls{ann} decoding with the same decoders used for mop-up for $Z$ basis memory experiments with uniform depolarisation parameter circuit noise. Results are shown for depolarisation parameters $p'=0.004$ and $p'=0.005$, corresponding to below-threshold performance.}
    \label{fig:higher-dist-timing}
\end{figure}

Decoder latency was tested by recording the time taken to calculate output corrections or logical corrections from the time the input data is prepared in an appropriate format. When performing these measurements, it was found that a significant dependence on input batch size was present, suggesting a significant overhead present in the benchmarking pipeline which does not grow with increasing batch size. To remove this initial overhead, which is not expected to be present in systems running syndrome decoding continuously during a large \gls{qec} benchmark, the time per shot was set as the linear dependence associated with a fit when batch sizes of 32, 64, 128, 192,and 256 were used. Results are shown in Figure~\ref{fig:ch5_timing} comparing \gls{pcm} PyMatching with unsimplified \gls{ann} decoding with \gls{pcm} PyMatching performing mop-up. We find that, for the test setup, composed of an Intel Core i9-13900K with an Nvidia GeForce RTX 4070 Ti, improvements to decoding times begin to appear only when above threshold and at code distances beyond $d=30$. This suggests that with moderate further optimisations, decoding speedups may be possible when utilising modest hardware. It should be noted that the time per shot, corresponding to the time needed to decode $n_c=d$ cycles, is still in the millisecond regime. Significant optimisations, perhaps with the use of dedicated hardware of \glspl{fpga} and \glspl{asic} may be needed to bring the total latency down to the microsecond regime. Results for larger distances are shown in Figure \ref{fig:higher-dist-timing}. The memory demands of larger distance simulations prohibit the higher batch-sizes necessary to find latencies by linear fits, and so these results show full times which include non-batch-dependent overheads. We find that crossover points do exist when \gls{ann}-based decoding is expected to be faster that \gls{pcm} and \gls{dem} PyMatching alone. Crossover points appear to occur at lower distances for higher error rates. We note an unexpected sensitivity of the ANN part of computations, especially at lower error rates, appears in the data. This may be an artefact of the testing process, as the \gls{ann} should be performing the same calculations irrespective of the mop-up component which would follow. 

\begin{figure}[htbp]
    \centering
    \includegraphics[width=\linewidth]{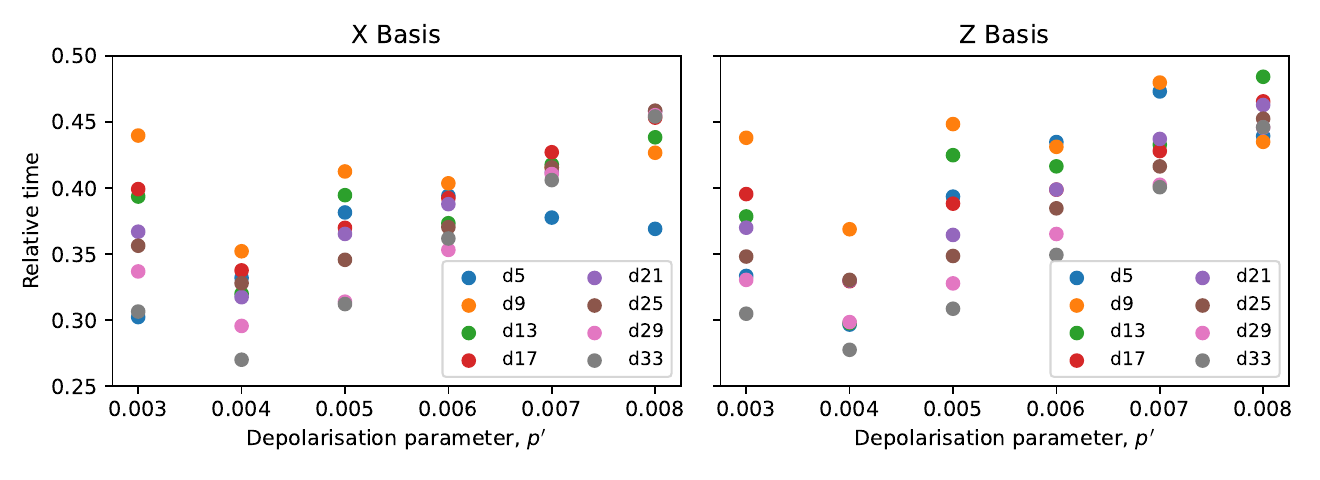}
    \caption{Relative matching decoding times for \gls{pcm} PyMatching decoding compared with multi-label \gls{ann} decoding with \gls{pcm} PyMatching for $X$ and $Z$ basis memory experiments with depolarisation parameter, $p'$, circuit noise.}
    \label{fig:ch5_relative_timing}
\end{figure}

Finally, we investigated the speedup associated with the global decoding task with and without the assistance of the \gls{ann}. Figure~\ref{fig:ch5_relative_timing} shows the relative time taken during matching when using \gls{pcm} PyMatching as a global decoder compared with \gls{ann} assistance. We find that relative time is consistently less than one, showing that a significant portion of the decoding problem is handled by the \gls{ann}. We find that relative time tends to increase for increasing noise intensities. We can also see that for each different noise model, relative decoding times tend to be lower for larger distances and greater for smaller distances. This suggests that significant advantages may be possible as long as sufficient optimisations are applied to make the highly parallel \gls{ann} operations contribute negligibly to the total decoding time.

\subsection{Comparison With Previous Works}
The most direct comparison of our work to previous research on \gls{ann} decoding at large distances is with Meinerz \etaldot~\cite{Meinerz2021arxiv} and Chamberland \etaldot~\cite{Chamberland_2023}. A summary is provided in Table \ref{review_table_meas}. Meinerz \etaldot used fully connected \glspl{ann} to form predictions for corrections near detection events on surface codes on a torus (toric codes). The application of dense networks near detection events has the advantage of requiring \gls{ann} computations only at a small subset of all correction locations, rather than for the whole syndrome history presented here. Considering surface codes defined on a torus has the advantage of perfect periodic boundary conditions, avoiding complications associated with boundaries. The error model considered was depolarising noise on data qubits with syndrome measurement errors on ancilla qubits. Meinerz \etaldot considered PyMatching \gls{mwpm} and \gls{uf} as global decoders, applied after the dense \gls{ann}. Timing measurements for \gls{ann}-assisted PyMatching decoding yielded times of $14.6$ ms and $229$ ms for error rates of $p=0.01$ and $p=0.0378$, respectively. Their implementation of PyMatching alone gave decoding latencies of $211$ ms and $294$ ms, consistent with the sparse syndrome targeted \gls{ann} decoding strategy. It should be noted that a slightly lower threshold was achieved when utilising the speedup of the \gls{ann} decoder before PyMatching. When applied with \gls{uf}, results showed the use of the \gls{ann} offered improvements to decoding times, lowering latencies from $11.5$ ms to $11.1$ ms at $p=0.01$ and $18.9$ ms to $17.8$ ms at $p=0.0378$. Although these offer significantly less improvements than when the \gls{ann} was paired with PyMatching, utilising the \gls{ann} with \gls{uf} yields an improved threshold from $p=0.0379$ to $p=0.0434$.

Chamberland \etaldot~\cite{Chamberland_2023} instead considered \gls{fcnn} decoding of rotated surface codes. The noise model considered was a variant of circuit noise, featuring only one instance of idle noise on data qubits each cycle. The \gls{ann} decoder under consideration was composed on multiple 3D convolution layers and was used to predict sites of data qubit errors based on syndrome measurement input data. As ancilla qubit errors were unable to be specified, heuristic methods were used to account for them before residual syndromes were processed by a global decoder. At low error rates of $p=0.001$, up to $98\%$ of nontrivial syndrome bits were able to be decoded, which was argued to suggest global decoder speedups by up to a factor of $10^6$. However, similar to the work of  Meinerz \etaldot~\cite{Meinerz2021arxiv}, when global decoding was performed with \gls{mwpm}, as implemented with PyMatching, threshold error rates decreased slightly from $p=0.007$ to $p=0.005$ when comparing PyMatching alone to using an \gls{ann}. Chamberland \etaldot also evaluated the latency of their \gls{ann} when implemented with \glspl{fpga} and found a total latency of $0.673$ ms for $17$ syndrome measurement rounds of a $d=17$ code.

\begin{table}[t]
\scriptsize
\centering
\caption{A comparison between this work and prior work on scalable \gls{ann}-based syndrome decoders for \gls{qec} codes with syndrome errors.}\label{review_table_meas}
\begin{tabular}{ |p{2.5cm}||p{2.0cm}|p{0.6cm}|p{0.6cm}|p{3.0 cm}|p{3.5cm}| }
 \hline
 \textbf{Paper}& \textbf{QEC Code} & \textbf{d$_{\mathrm{max} }$} &\textbf{p$_{\mathrm{th} }$} & \textbf{ML Technique} & \textbf{Noise model }\\
 \hline
 
 Meinerz \etaldot \newline(2018)~\cite{Meinerz2021arxiv} & Toric code  & 63 & 0.044 &  Fully connected \gls{ann} & Phenomenological noise\\
 \hline
 
 Chamberland \etaldot \newline(2018)~\cite{Chamberland_2023} & Rotated surface code  & 17 &0.005 & \gls{fcnn} & Circuit noise (uniform error rate)\\
 \hline
 \hline
 This work. & Rotated surface code  & 97 &  0.006 & \gls{fcnn}, Diffusion model &  Circuit noise (uniform depolarisation parameter) \\
\hline

\end{tabular}
\end{table}

\subsection{Comparison With Currently Available Devices}\label{subsec:compare_sc_ion}
Varying characteristics of hardware platforms supporting \gls{qec} impose different requirements for the surface code decoding problem. The major differences derive from qubit and gate error characteristics, two-qubit connectivity, gate times and classical resource availability. Here, we will give a brief discussion of the characteristics of currently available superconducting and trapped atom devices to put in context the results presented in this work.

Superconducting quantum devices, such as those developed by IBM and Google~\cite{Gupta_2024, acharya2022suppressing}, possess amongst the fastest gate times, with two qubit gates possible at tens of nanoseconds and measurements at hundreds of nanoseconds. These devices are expected to output syndrome data at $1 \mu\mathrm{s}$ per cycle. The error rates of the best available superconducting quantum devices are currently overall slightly below threshold, but feature significant inhomogeneity between contributions from two-qubit gates, single qubit gates, and \gls{spam} errors. Moreover, these devices are known to experience significant contributions from other error sources, such as cross-talk, leakage and cosmic ray events \cite{chen2021nature}. Such devices have grown to sizes featuring beyond a hundred qubits, and have demonstrated surface code logical qubits of distances up to $d=7$ \cite{acharya2024}. Connectivity to control electronics is expected to be a significant challenge as such devices continue to scale to sizes featuring multiple logical qubits of distances up to approximately $d=33$. Such overheads are expected to be required for the largest algorithms under pessimistic estimates for achievable physical error rates of approximately $p=0.1\%$. In the near future, when devices grow to sizes corresponding to distances supporting beyond $d=13$, utilising 3D convolutional \glspl{ann} may offer some speed-ups. However, this would require latencies of hardware implementations of \glspl{ann} to be negligible compared to the global decoder latency. Additionally, as the noise of such devices can be expected to remain significantly structured, additional research investigating adaptations to structured noise models may be needed to maintain competitive performance compared to weighted matching methods alone.

Trapped atom/ion devices, such as those developed by Quantinuum and Quera, possess much longer qubit coherence times, but also suffer from much longer gate times~\cite{Evered2023, mayer2024benchmarkinglogicalthreequbitquantum}. A major advantage of such devices compared to other platforms comes from the all-to-all connectivity possible from long range interactions and movement of atomic qubits in optical lattices. The increased connectivity compared to local architectures can be used to implement transversal two-qubit logical operations. When paired with simultaneous decoding of correlated errors, and used of optimised compiling, this has the advantage of overcoming the negative impact of slower gate times~\cite{litinski2022activevolumearchitectureefficient, cain2025correlateddecodinglogicalalgorithms}. The convolutional decoder presented here has not been optimised for the simultaneous decoding of multiple logical qubits with correlated errors. However, single qubit decoding during idle times may benefit from convolutional \gls{ann} decoding speedups, especially if large delays are caused by optimal correlated decoding after transversal logical operations.

\section{Discussion and Outlook}\label{sec:Discussion}
In this work we investigated a 3D convolutional approach to \gls{ann}-based, circuit noise surface code decoding. We described a vectorised method of data preparation, corresponding to parallel propagation of errors to reference time steps and how decoding can be interpreted as a multi-label classification and generative modelling problem. We showed how this allows an \gls{ann} based on three-dimensional convolutional layers to be trained on such a task, and evaluated performance with respect to accuracy and latency. We find that such a decoder exhibits competitive performance which may be valuable addition to experimental \gls{qec} architectures performing the decoding required for \gls{ftqc}.

There are many ways our work can be extended. On the decoding side, further software optimisations, such as different unit cells, quantization, pruning and parallelisation may further improve desktop performance. Additionally, optimisation can be applied to generative approaches to decoding, such as utilizing noise schedules for diffusion models or avoiding the need to repeated passes by using \glspl{gan} instead. Further improvements to performance may also be sought when implemented on dedicated hardware. A complete evaluation when implemented with dedicated hardware would also provide strong information regarding the presence of any bottlenecks present during computation, such as data transfer. Our network was shown to operate only on uniform surface codes, where stabilisers form a checker-board pattern. Some lattice surgery configurations, which are needed to apply a universal set of logical operators, can also make use of domain walls and twists. An implementation with modified unit cell may be able to accommodate such codes, with generalisation expected to follow if layers remain exclusively convolutional. Additionally, the network can be optimised to work with data when received as a data stream, a possibility readily enabled by the local computational structure in space and time. Once this is achieved, decoding general lattice surgery operations, changing configuration with time, may be possible. On the simulation side, further work can be done to extend the results to other periodic \gls{qec} codes such as colour codes.

In summary, the techniques developed in this work show that neural-network based decoding can remain competitive and scalable up to code distances relevant to those expected to yield quantum advantage. With further development, such methods form candidates to assist in the classical data processing demanded by \gls{qec} in practical settings.

\section{Data availability statement}
The data that support the findings of this study are available on request from the corresponding author upon reasonable request.

\section{Acknowledgements}
This work was supported by the Australian Research Council funded Center for Quantum Computation and Communication Technology (CE170100012).

\section{References}

\bibliographystyle{quantum} 









\end{document}